\documentclass[a4paper,twoside,twocolumn,english,aps,superscriptaddress,showpacs]{revtex4}
\usepackage{times}
\usepackage[T1]{fontenc}
\usepackage[latin1]{inputenc}
\usepackage{subfigure}
\usepackage{amsmath}
\usepackage{graphicx}
\usepackage{amssymb}

\makeatletter

\newcommand{\lyxdot}{.}

\usepackage{babel}
\makeatother
\begin{document}

\title{Binding between two-component bosons in one dimension}

\author{Emmerich Tempfli}

\affiliation{Theoretische Chemie, Universit\"{a}t Heidelberg, Im Neuenheimer
Feld 229, 69120 Heidelberg, Germany}

\author{Sascha Zöllner}

\email{sascha.zoellner@pci.uni-heidelberg.de}

\affiliation{Theoretische Chemie, Universit\"{a}t Heidelberg, Im Neuenheimer
Feld 229, 69120 Heidelberg, Germany}

\author{Peter Schmelcher}

\email{peter.schmelcher@pci.uni-heidelberg.de}

\affiliation{Theoretische Chemie, Universit\"{a}t Heidelberg, Im Neuenheimer
Feld 229, 69120 Heidelberg, Germany}

\affiliation{Physikalisches Institut, Universit\"{a}t Heidelberg, Philosophenweg
12, 69120 Heidelberg, Germany}

\pacs{03.75.Mn, 67.60.Bc, 67.85.-d}

\date{May 7, 2009}

\begin{abstract}
We investigate the ground state of one-dimensional few-atom Bose-Bose
mixtures under harmonic confinement throughout the crossover from
weak to strong inter-species \textit{attraction}. The calculations
are based on the numerically exact multi-configurational time-dependent
Hartree method. For repulsive components we detail the condition for
the formation of a molecular Tonks-Girardeau gas in the regime of
intermediate inter-species interactions, and the formation of a molecular
condensate for stronger coupling. Beyond a critical inter-species
attraction, the system collapses to an overall bound state. Different
pathways emerge for unequal particle numbers and intra-species interactions.
In particular, for mixtures with one attractive component, this species
can be viewed as an effective potential dimple in the trap center
for the other, repulsive component.
\end{abstract}
\maketitle

\section{Introduction}

Cold atoms have become an important tool to create and study strongly
correlated quantum systems \citep{bloch07,lewenstein07}. One main
reason is that it is possible to experimentally tune the effective
low-energy interaction strength between the atoms using Feshbach resonances
\citep{koehler06}. This has proven useful particularly for Fermi
gases \citep{giorgini07}, whereas for bosons the creation of strong
interactions is limited by three-body collisions. However, in lower
(here: one) dimensions there are also other possibilities of achieving
effectively strong correlations -- e.g., by lowering the atom-number
density \citep{lieb63a} and via confinement-induced resonances, which
exploit the parametric dependence on the transverse trapping potential
\citep{Olshanii1998a}. This allows one to practically adjust the
coupling strength all the way from infinite attraction to hard-core
repulsion.

For \emph{single-component} bosons in one dimension (1D), the extreme
case of infinitely repulsive interactions is known as the Tonks-Girardeau
gas, which has been realized experimentally \citep{kinoshita04,paredes04}.
Here the system maps to an ideal gas of fermions, in the sense that
the exclusion principle emulates the effect of hard-core \emph{}repulsion
\citep{girardeau60}. The microscopic mechanism of the crossover from
the weakly interacting Bose gas to the above \emph{fermionization}
limit has been investigated in detail \citep{alon05,hao06,zoellner06a,zoellner06b,deuretzbacher06,schmidt07}.
By contrast, the ground state for strong \emph{attraction} is an $N$-atom
molecule \citep{mcguire64}. However, exotic fermionized excitations
exist for sufficiently attractive interactions \citep{astrakharchik05,batchelor05,tempfli08}.

In the case of \emph{two} (or more) bosonic components\emph{,} a plethora
of configurations exists: On top of varying both intra- and inter-species
interactions, also the trapping potentials may be made species dependent.
Moreover, the experimental availability of different two-component
mixtures (involving not only different hyperfine components \citep{myatt97,stenger98}
or isotopes \citep{papp08,fukuhara09}, but altogether different atomic
species like K-Rb \citep{modugno02}) adds another degree of flexibility.
For two 1D Bose gases with inter-species \emph{repulsion}, a generalized,
composite fermionization exists which may lead to demixing of the
two components atom by atom \citep{girardeau07,zoellner08b,hao08,hao09}.
In a lattice potential, even more complex patterns have been found,
cf. \citep{alon06,mishra07,roscilde07,kleine08} and Refs. therein.

In this work, we are interested in the \emph{binding} between two
bosonic species, i.e., the crossover from weak to strong inter-species
\emph{attraction}. Here little is known except for a general classification
based on the harmonic-fluid approximation \citep{cazalilla03}. For
\emph{fermions}, pairing between the two components has been predicted,
which then form a Tonks-Girardeau gas of molecules \citep{astrakharchik04b}.
Similarly, pairing has been found in attractive Bose-Fermi mixtures
in various settings \citep{salasnich07,pollet08,rizzi08}. While,
in principle, a 1D fermionic component maps to a strictly fermionized
bosonic one, the physics of realistic Bose-Bose mixtures differs in
two ways: For one thing, the finite intra-species repulsion must compete
with strong inter-species attraction. More generally, in contrast
to fermions all possible \emph{intra}-species interactions are possible
and make for interesting phases. The key goal of this work is to demonstrate
effects due to the interplay between intra-species and inter-species
forces.

Our paper is organized as follows. Section~\ref{sec:method} introduces
the model and briefly reviews the computational method. The pairing
between repulsive components is elucidated in Sec.~\ref{sec:Repulsive},
first for the case of a mixture of balanced components (Secs.~\ref{sub:RepulsiveA},
\ref{sub:RepulsiveB}), complemented by a discussion of atom-number
imbalances and unequal intra-species interactions (Sec.~\ref{sub:RepulsiveC}).
Section~\ref{sec:Attractive} deals with the question of how the
presence of attractive components alters the picture.

\section{Model and computational method\label{sec:method}}

\paragraph*{Model}

The object of investigation is a two-component Bose gas (denoted $\alpha\in\left\{ A,B\right\} $)
subjected to a one-dimensional confinement, where the external potentials
acting on the different species are assumed to be the same, i.e. $U_{\alpha}\left(x\right)=U\left(x\right)$.
The two species can be considered as two internal states (pseudospin
$\left|\uparrow\right\rangle $ and $\left|\downarrow\right\rangle $)
of the same kind of Bose atoms, or as different isotopes with the
mass $m_{\alpha}\approx m$. In the subsequent sections we denote
the atom number of each species with $N_{\alpha}$ and the total number
with $N=N_{A}+N_{B}$. This kinematically one-dimensional system of
trapped bosons can be described in the low-energy limit by an effective
one-dimensional Hamiltonian with contact interactions. The second
quantized Hamiltonian $H$ then reads \begin{equation}
\begin{array}{ccl}
H & = & \int dx\underset{\alpha=A,B}{\sum}\left\{ \hat{\Psi}_{\alpha}^{\dagger}\left(x\right)\left[-\frac{1}{2}\frac{\partial^{2}}{\partial x^{2}}+U\left(x\right)\right]\hat{\Psi}_{\alpha}\left(x\right)\right.\\
 &  & +\left.\frac{g_{\alpha}}{2}\hat{\Psi}_{\alpha}^{\dagger}\left(x\right)\hat{\Psi}_{\alpha}^{\dagger}\left(x\right)\hat{\Psi}_{\alpha}\left(x\right)\hat{\Psi}_{\alpha}\left(x\right)\right\} \\
 &  & +g_{AB}\int dx\hat{\Psi}_{A}^{\dagger}\left(x\right)\hat{\Psi}_{B}^{\dagger}\left(x\right)\hat{\Psi}_{B}\left(x\right)\hat{\Psi}_{A}\left(x\right),\end{array}\label{eq:0.1}\end{equation}
where the field operator $\hat{\Psi}_{\alpha}\left(x\right)$ ($\hat{\Psi}_{\alpha}^{\dagger}\left(x\right)$)
annihilates (creates) a boson of the $\alpha$-species at the position
$x$. The effective intra- and inter-species couplings $g_{\alpha}$
and $g_{AB}$ characterize the interaction between the atoms and can
be controlled experimentally by the scattering lengths $a_{0}^{\left(\alpha\right)}$
and $a_{0}^{\left(AB\right)}$, respectively, in analogy to the single
species case \citep{Olshanii1998a} . Furthermore, the standard rescaling
procedure to harmonic oscillator units has been carried out (cf. \citep{zoellner08b}
for details). For technical reasons we apply the Hamiltonian in the
first quantized form. The eigenvalue problem reduces to solving the
stationary Schrödinger equation $H\Psi=E\Psi$, with $H\equiv\sum_{\alpha}H_{\alpha}+H_{AB}$
composed of the single species Hamiltonian 

\[
\begin{array}{ccc}
H_{\alpha} & = & \sum_{i=1}^{N_{\alpha}}\left[\frac{1}{2}p_{\alpha_{i}}^{2}+U(x_{\alpha_{i}})\right]+\underset{i<j}{\sum}g_{\alpha}\delta_{\sigma}(x_{\alpha_{i}}-x_{\alpha_{j}})\end{array}\]
and the inter-species coupling part \[
H_{\mathrm{AB}}=\sum_{i=1}^{N_{A}}\sum_{j=1}^{N_{B}}g_{\mathrm{AB}}\delta_{\sigma}(x_{\mathrm{A}_{i}}-x_{B_{j}}).\]
Here the effective 1D contact interaction potential is mollified with
a Gaussian $\delta_{\sigma}(x)\equiv e^{-x^{2}/2\sigma^{2}}/\sqrt{2\pi}\sigma$
(of width $\sigma=0.05$) for numerical reasons. In the further examination
we focus on the case of a harmonic confinement, $U(x)=\frac{1}{2}x^{2}$
and on attractive inter-species forces $g_{AB}\in(-\infty,0]$. (The
case of repulsive inter-species couplings has already been investigated
in \citep{zoellner08b}. Note that, in the case of $U=0$ and $g_{\alpha}=g_{AB}$,
this system is integrable via Bethe's ansatz as in \citep{lieb63a}.)\\

\paragraph*{Computational method }

Our approach relies on the numerically exact multi-configuration time-dependent
Hartree method \citep{mey90:73,bec00:1,mctdh:package}, a quantum-dynamics
approach which has been applied successfully to systems of few identical
bosons \citep{zoellner06a,zoellner06b,zoellner07a,zoellner07c,tempfli08,eckart09}
as well as to Bose-Bose mixtures \citep{zoellner08b}. Its principal
idea is to solve the time-dependent Schrödinger equation $\begin{array}{c}
i\dot{\Psi}(t)=H\Psi(t)\end{array}$ as an initial-value problem by expanding the solution in terms of
direct (or Hartree) products $\Phi_{J}\equiv\varphi_{j_{1}}^{\left(1\right)}\otimes\cdots\otimes\varphi_{j_{N}}^{\left(N\right)}$:\begin{equation}
\Psi(t)=\sum_{J}A_{J}(t)\Phi_{J}(t).\label{eq:mctdh-ansatz}\end{equation}
The unknown single-particle functions $\varphi_{j}^{\left(\kappa\right)}$
($j=1,\dots,n_{\kappa}$) are in turn represented in a fixed \emph{}basis
of, in our case, harmonic-oscillator orbitals. The specific feature
of the system at hand is the indistinguishability within each species.
Therefore the single-particle functions are identical within each
subset $K_{A}=\left\{ 1,\ldots,N_{A}\right\} $ and $K_{B}=\left\{ N_{A}+1,\ldots,N\right\} $
(, i.e. $\varphi_{j}^{\left(\kappa\right)}=\varphi_{j}^{\left(\alpha\right)},\,\forall\kappa\in K_{\alpha}$).
The permutation symmetry within each subset is ensured by the correct
symmetrization of expansion coefficients $A_{J}$. In analogy to the
wave function, also non-separable terms of the Hamiltonian such as
the two-body interaction are expanded in terms of direct products
\citep{jae96:7974}.

Note that, in the above expansion, not only the coefficients $A_{J}$
but also the single-particle functions $\varphi_{j}$ are time dependent.
Using the Dirac-Frenkel variational principle, one can derive equations
of motion for both $A_{J},\varphi_{j}$ \citep{bec00:1}. Integrating
this differential-equation system allows us to obtain the time evolution
of the system via (\ref{eq:mctdh-ansatz}). This has the advantage
that the basis set $\{\Phi_{J}(t)\}$ is variationally optimal at
each time $t$; thus it can be kept relatively small. Still, its exponential
growth with the number of particles limits our approach to only a
few atoms $N<10$, depending on how many single-particle functions
need to be included to describe inter-particle correlations.

Although designed for time-dependent studies, it is also possible
to apply this approach to stationary states. This is done via the
so-called \emph{relaxation} method. The key idea is to propagate some
wave function $\Psi(0)$ by the non-unitary $e^{-H\tau}$ (propagation
in imaginary time.) As $\tau\to\infty$, this exponentially damps
out any contribution but that originating from the true ground state
like $e^{-(E_{m}-E_{0})\tau}$. In practice, one relies on a more
sophisticated scheme termed \emph{improved relaxation}, which is much
more robust especially for excitations. Here $\langle\Psi|H|\Psi\rangle$
is minimized with respect to both the coefficients $A_{J}$ and the
orbitals $\varphi_{j}$. The effective eigenvalue problems thus obtained
are then solved iteratively by first solving for $A_{J}$ with \emph{fixed}
orbitals and then {}`optimizing' $\varphi_{j}$ by propagating them
in imaginary time over a short period. That cycle will then be repeated.

\section{Mixture of two repulsive components\label{sec:Repulsive}}

In this section we investigate two repulsive components ($g_{\alpha}>0$)
with increasing inter-species attraction $g_{AB}\in\left(-\infty,0\right]$.
We start with components of equal intra-species settings, such as
equal intra-species interaction strengths, $g_{A}=g_{B}$, and particle
numbers $N_{A}=N_{B}$ and discuss subsequently the changes in the
system's behavior when relaxing these conditions.

\subsection{Mixture of two fermionized components\label{sub:RepulsiveA}}

The starting point is the highly repulsive limit of the components,
i.e. two quasi-fermionized states within the two species with the
inter-species interactions $g_{A}=g_{B}=25.0$. For small inter-species
attraction $g_{AB}=-0.001$ the system is well described by the uncorrelated
product of two \emph{Tonks-Girardeau} (TG) states $\Psi=\Psi_{A}\otimes\Psi_{B}$,
where $\Psi_{A}=\Psi_{B}\approx\left|\Psi_{0}^{F}\right|$. This means,
the state of each species $\alpha$ in the high-interaction limit
$\left(g_{\alpha}\rightarrow+\infty\right)$ can be mapped to a non-interacting
state $\Psi_{0}^{F}$ of identical fermions \citep{girardeau60},
also commonly termed \textit{fermionization}. By extension, a mixture
of two fermionized Bose gases has similarities with a two component
Fermi gas.

\begin{figure}
\includegraphics[width=0.7\columnwidth,keepaspectratio]{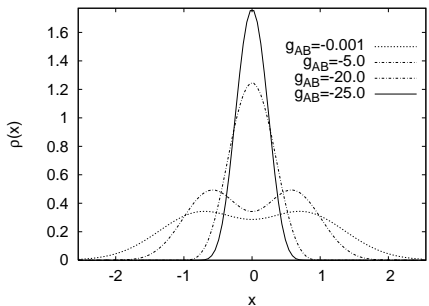}

\caption{One-body density $\rho\left(x\right)\equiv\rho^{\left(\alpha\right)}\left(x\right)$
($\alpha\in\left\{ A,B\right\} $) for a quasi-fermionized mixture
($g_{\sigma}=25.0$) with the particle numbers $N_{\sigma}=2$, plotted
for different inter-species interactions $g_{AB}$ (see legend). \label{fig:Pair1}}
\end{figure}

The characteristic fermionic pattern of the TG-state is displayed
in the one-body density (which is the same for both species $\alpha$
for symmetry reasons) $\rho\left(x\right)=\rho^{\left(\alpha\right)}\left(x\right)$
(with $\alpha\in\left\{ A,B\right\} $), which measures the probability
distribution of finding one $\alpha$ particle at the position $x$
pictured in Fig. \ref{fig:Pair1}. %
\begin{figure}
\includegraphics[width=0.35\columnwidth]{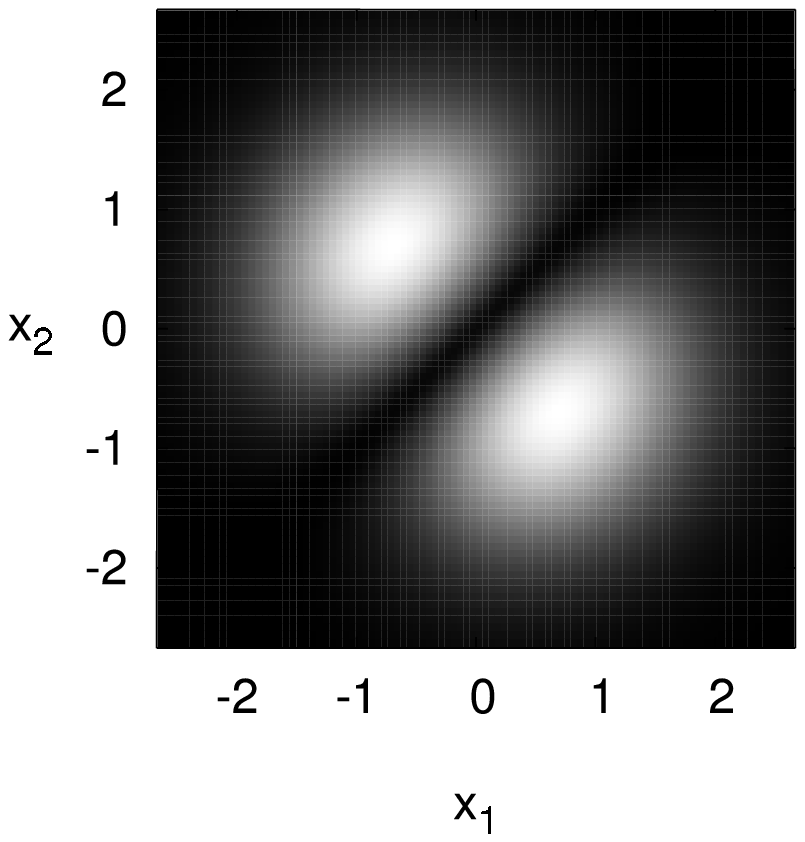}~~~~~~~~~~~~~\includegraphics[width=0.35\columnwidth]{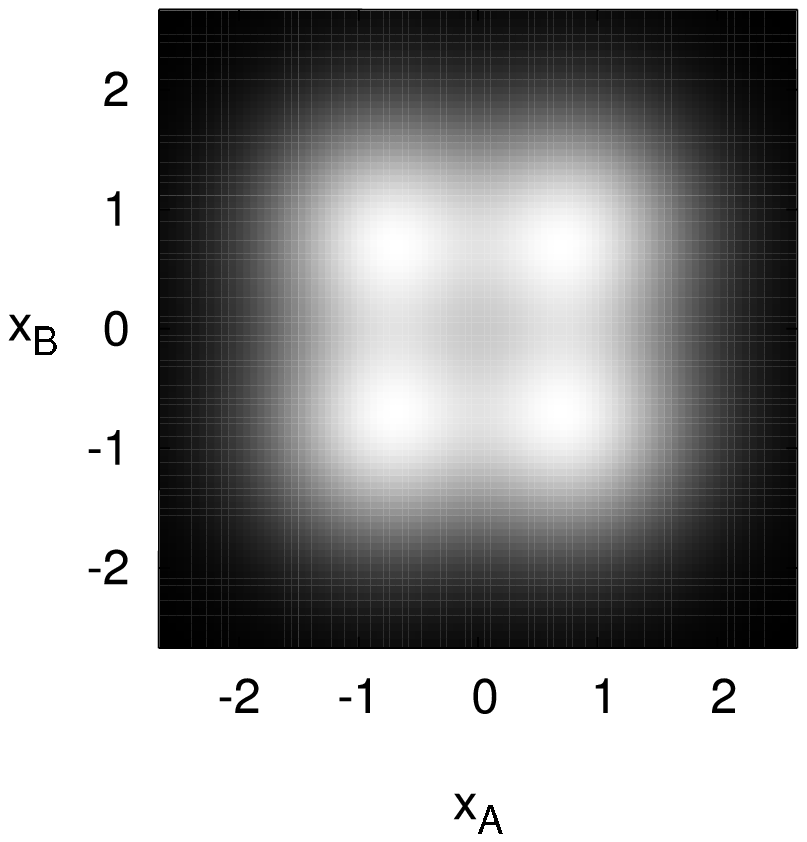}

\includegraphics[width=0.35\columnwidth]{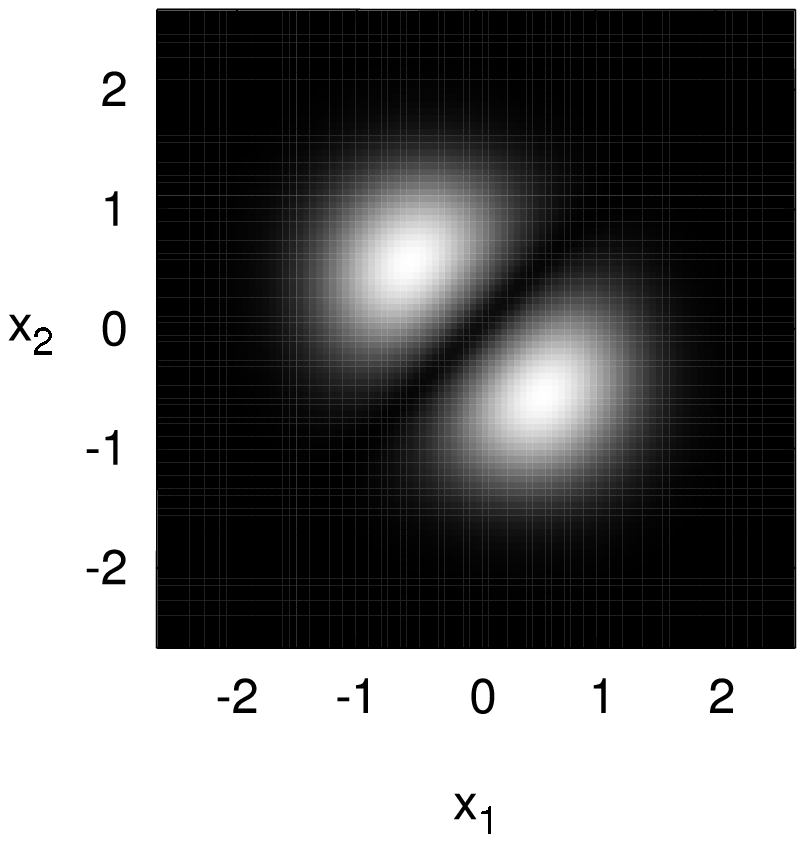}~~~~~~~~~~~~~\includegraphics[width=0.35\columnwidth,keepaspectratio]{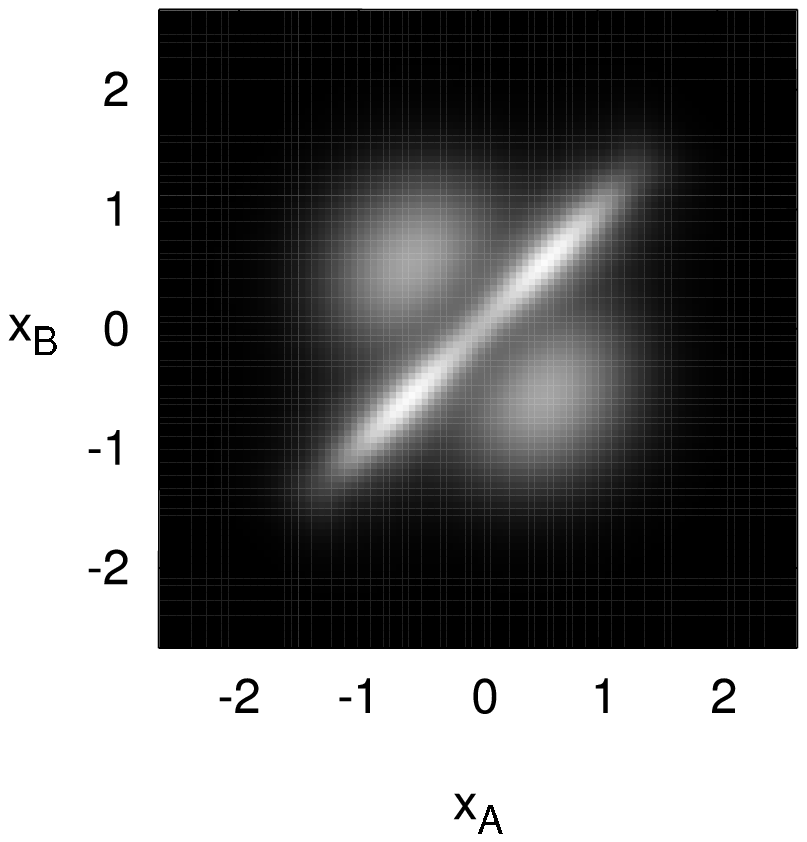}

\includegraphics[width=0.35\columnwidth]{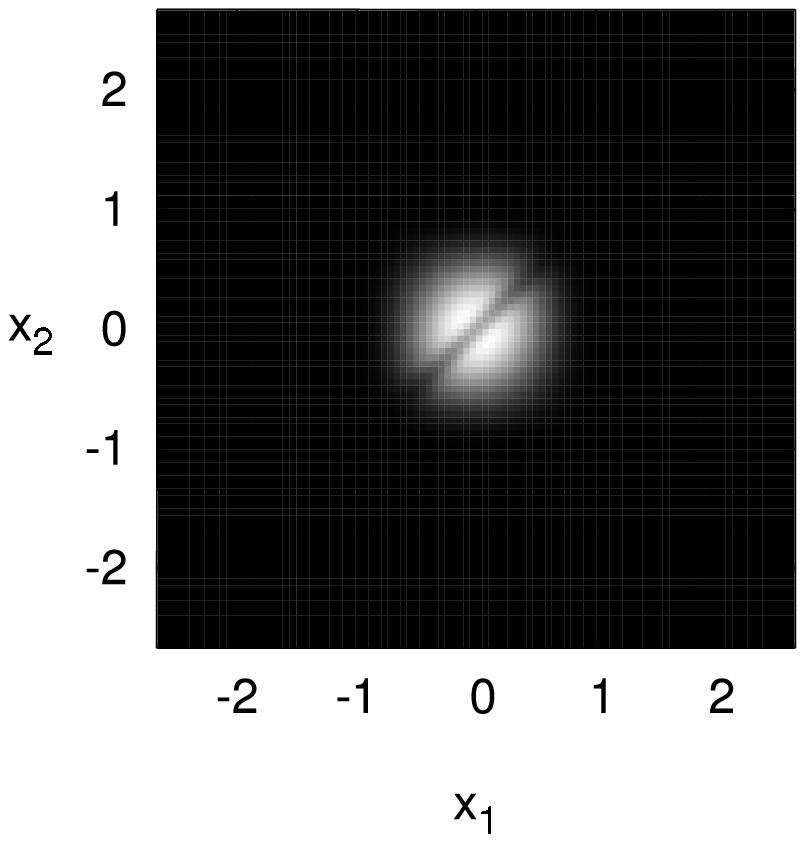}~~~~~~~~~~~~~\includegraphics[width=0.35\columnwidth]{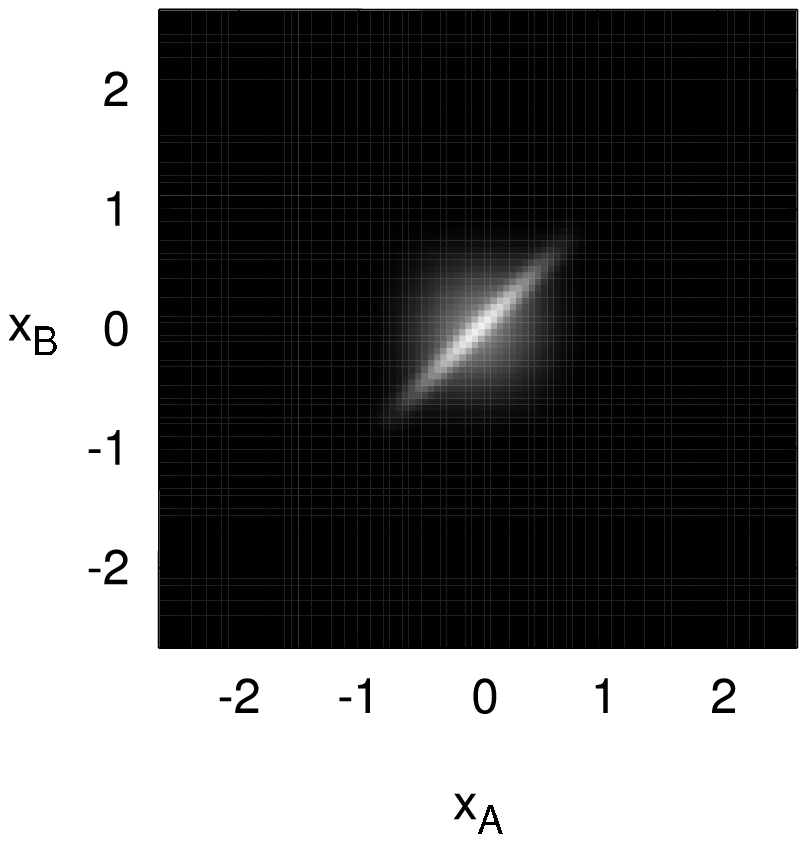}

\includegraphics[width=0.35\columnwidth]{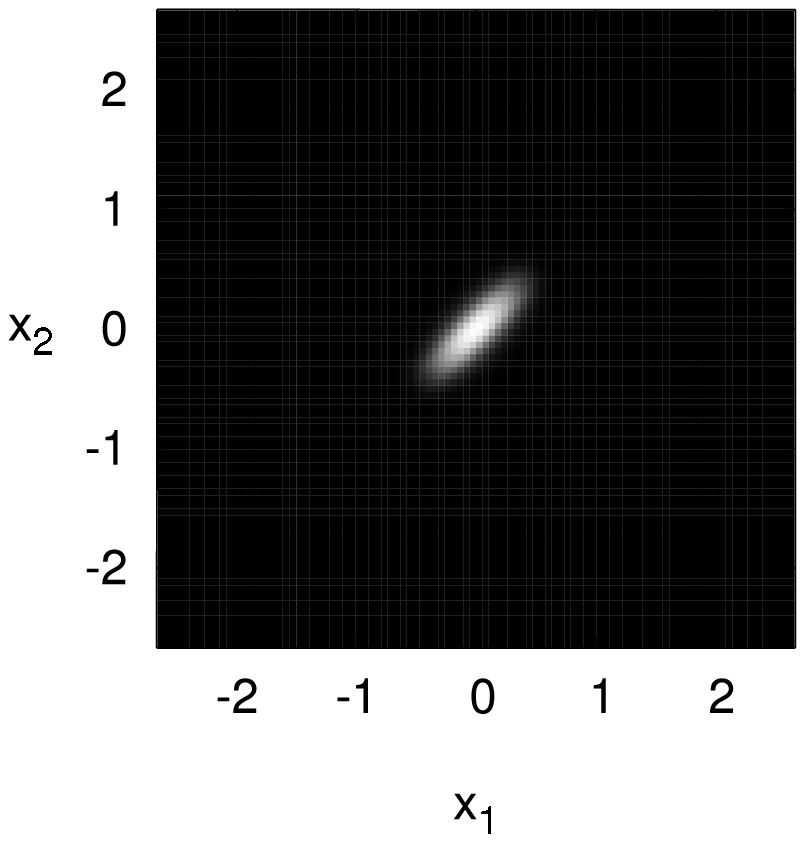}~~~~~~~~~~~~~\includegraphics[width=0.35\columnwidth]{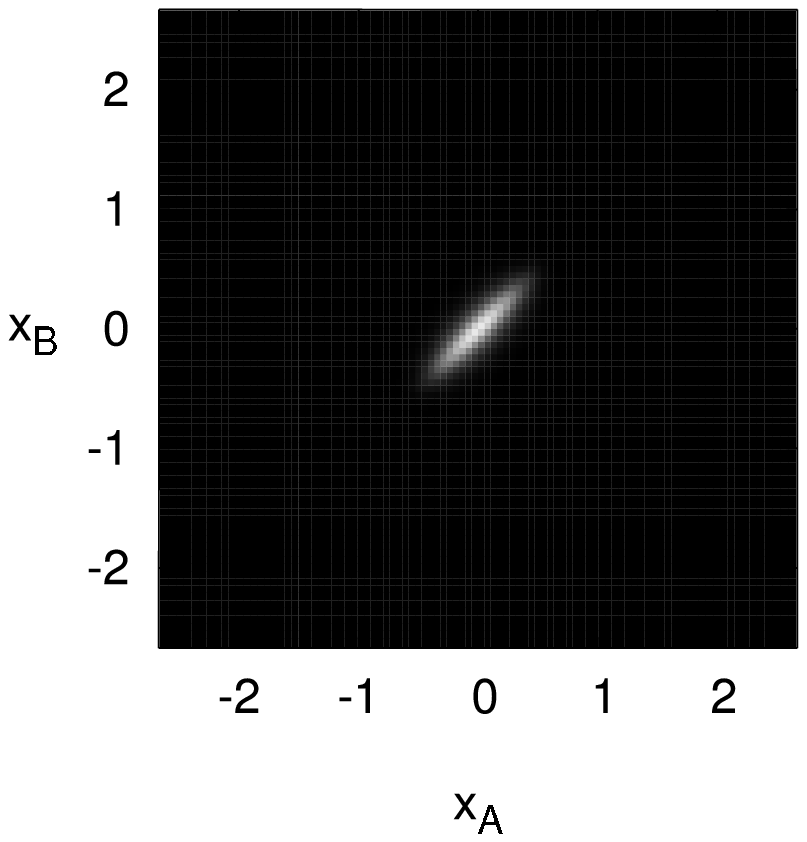}

\caption{Two-body correlation functions $\rho_{\alpha\alpha}\left(x_{1},x_{2}\right)$
\textit{(left column}) and $\rho_{AB}\left(x_{A},x_{B}\right)$ (\textit{right
column}) of two quasi-fermionized components $g_{A}=g_{B}=25.0$ for
inter-species couplings $g_{AB}=-0.001,\,-5.0,\,-20.0,\,-30.0$ (\textit{from
top to bottom}).\label{fig:Pair2}}
\end{figure}
 One recognizes the density concentrating more and more in the center
of the trap with increasing inter-species attraction $g_{AB}$. Concurrently,
in the intermediate interaction regime $\left|g_{AB}\right|=5$ the
initial two density peaks are at first getting increasingly pronounced,
whereas the density in the center of the trap grows slowly. That intensification
of the fermionic characteristic is due to molecule formation, as discussed
in the following. By contrast, in the very high interaction regime
$\left(\left|g_{AB}\right|=20\sim g_{\alpha}\right)$ the two peaks
merge into one single peak in the center of the trap.

A more detailed insight in the systems behavior is given by the two-body
correlation functions. In the case of a binary mixture, these are
defined as\[
\begin{array}{l}
\rho_{\alpha\alpha}\left(x_{1},x_{2}\right)=\frac{1}{N_{\alpha}\left(N_{\alpha}-1\right)}\hat{\left\langle \Psi\right.}_{\alpha}^{\dagger}\left(x_{1}\right)\hat{\Psi}_{\alpha}^{\dagger}\left(x_{2}\right)\hat{\Psi}_{\alpha}\left(x_{2}\right)\hat{\Psi}_{\alpha}\left.\left(x_{1}\right)\right\rangle \\
\rho_{AB}\left(x_{A},x_{B}\right)=\frac{1}{N_{A}N_{B}}\hat{\left\langle \Psi\right.}_{A}^{\dagger}\left(x_{A}\right)\hat{\Psi}_{B}^{\dagger}\left(x_{B}\right)\hat{\Psi}_{B}\left(x_{B}\right)\hat{\Psi}_{A}\left.\left(x_{A}\right)\right\rangle .\end{array}\]
The two-body correlation functions $\rho_{\alpha\alpha}\left(x_{1},x_{2}\right)$
and $\rho_{AB}\left(x_{A},x_{B}\right)$ in Fig. \ref{fig:Pair2},
depict the conditional probability of measuring the one $\alpha$-particle
at the position $x_{1}$ and the other at the position $x_{2}$, and
likewise for the different species, $\rho_{AB}\left(x_{A},x_{B}\right)$.
The left column in Fig. \ref{fig:Pair2} indicates that the two species
keep their {}``fermionic'' character, i.e., the probability of finding
particles of the same kind at the same position $\left\{ x_{1}=x_{2}\right\} $
stays very low up to high inter-species attractions $\left(\left|g_{AB}\right|\lesssim20\right)$.
But according to $\rho_{AB}\left(x_{A},x_{B}\right)$ (right column),
the two species concentrate more and more at the same position, which
means on the diagonal in two separate peaks aside the center of the
harmonic trap (see Fig. \ref{fig:Pair2}, $\rho_{AB}\left(x_{A},x_{B}\right)$
for $g_{AB}=-5.0$). This can be understood as formation of a \textit{molecular
Tonks-Girardeau} (MTG) state: As we shall argue below, two distinguishable
particles form a bound state, that is a molecule (in the following
denoted as $AB$-molecule). These indistinguishable $AB$-molecules
in turn form a (molecular) Tonks-Girardeau state. Whereas for two-component
Fermi gases this MTG state remains stable even in the strongly attractive
inter-species attraction regime \citep{astrakharchik04b}, this is
not the case in a pure bosonic mixture (see also Fig. \ref{fig:Pair2}).

\subsubsection*{Pairing description}

For a better understanding of this behavior we examine the Hamiltonian
for the exemplary case $N_{\alpha}=2$. To this end, we transform
$X\equiv\left(x_{A_{1}},x_{A_{2}};x_{B_{1}},x_{B_{2}}\right)^{\top}$
to the relative coordinates $Y=\left(R_{CM},R_{1},r_{1},r_{2}\right)^{\top}$specified
by

\begin{equation}
Y=\mathcal{O}X,\,\,\mathcal{O}=\left(\begin{array}{cccc}
\frac{1}{\sqrt{4}} & \frac{1}{\sqrt{4}} & \frac{1}{\sqrt{4}} & \frac{1}{\sqrt{4}}\\
\frac{1}{2} & -\frac{1}{2} & \frac{1}{2} & -\frac{1}{2}\\
\frac{1}{\sqrt{2}} & 0 & -\frac{1}{\sqrt{2}} & 0\\
0 & \frac{1}{\sqrt{2}} & 0 & -\frac{1}{\sqrt{2}}\end{array}\right).\label{eq:Pair1}\end{equation}
The coordinates $R_{CM}$ ($r_{1,},r_{2}$) coincide - up to a factor
- with the standard center-of-mass (inter-species relative) coordinates.
The coordinate $R_{1}=\frac{1}{2}\left[\left(x_{A_{1}}+x_{B_{1}}\right)-\left(x_{A_{2}}+x_{B_{2}}\right)\right]$
specifies the distance between the centers of mass of two $\left(A,B\right)$-clusters.
The orthogonal transformation leads to the Hamiltonian $H\left(Y\right)=h_{CM}\left(R_{CM}\right)+H_{rel}$,
with

\begin{equation}
\begin{array}{l}
H_{rel}=\left[\frac{1}{2}p_{R_{1}}^{2}+\frac{1}{2}R_{1}^{2}\right]+\underset{i=1}{\overset{2}{\sum}}\left[\frac{1}{2}p_{r_{i}}^{2}+\frac{1}{2}r_{i}^{2}+\frac{g_{AB}}{\sqrt{2}}\delta\left(r_{i}\right)\right]\\
\,\,\,\,\,+g_{A}\delta\left(\frac{1}{\sqrt{2}}\left(r_{1}-r_{2}\right)-R_{1}\right)+g_{B}\delta\left(\frac{1}{\sqrt{2}}\left(r_{1}-r_{2}\right)+R_{1}\right)\\
\,\,\,\,\,+g_{AB}\underset{\pm}{\sum}\delta\left(\frac{1}{\sqrt{2}}\left(r_{1}+r_{2}\right)\pm R_{1}\right).\end{array}\label{eq:Pair2}\end{equation}
If we assume the formation of, say, $A_{i}B_{i}$-bound states ($i\in\left\{ 1,2\right\} $)
(up to permutation symmetry), for high enough inter-species attraction
$g_{AB}$ the extension of an $A_{i}B_{i}$- molecule is much smaller
than the distances between two such molecules $\left(\left|r_{i}\right|\ll\left|R_{1}\right|\right)$.
One can check this in the two-body correlation functions (Fig. \ref{fig:Pair2}).
In this limit, we can approximate \eqref{eq:Pair2} by the decoupled
Hamiltonian

\begin{equation}
\begin{array}{ccl}
H_{rel} & \approx & \underset{i=1}{\overset{2}{\sum}}\left[\frac{1}{2}p_{r_{i}}^{2}+\frac{1}{2}r_{i}^{2}+\frac{g_{AB}}{\sqrt{2}}\delta\left(r_{i}\right)\right]\\
 &  & +\left[\frac{1}{2}p_{R_{1}}^{2}+\frac{1}{2}R_{1}^{2}+\tilde{g}\delta\left(R_{1}\right)\right],\end{array}\label{eq:Pair3}\end{equation}
the last part describing the relative motion of the two $AB$-molecules
with the effective interaction $\tilde{g}\equiv g_{A}+g_{B}+2g_{AB}$.
The analytic solution of the ground state is known \citep{Busch98}
and the relative part (excluding the trivial CM factor) can be written
as \citep{tempfli08}

\begin{equation}
\begin{array}{l}
\psi_{rel}\left(X\right)\propto S_{+}\left\{ \left(\underset{i=1,2}{\prod}e^{-\frac{\left|g_{AB}\right|}{2}\left|x_{A_{i}}-x_{B_{i}}\right|}\right)\right.\\
\,\,\,\,\,\,\,\,\,\,\,\left.\times U\left(-\epsilon\left(\tilde{g}\right),\frac{1}{2}\left[\left(x_{A_{1}}+x_{B_{1}}\right)-\left(x_{A_{2}}+x_{B_{2}}\right)\right]\right)\right\} ,\end{array}\label{eq:Pair4}\end{equation}
where $\epsilon\left(\tilde{g}\right)=\nu\left(\tilde{g}\right)+1/2$
is determined by the transcendental equation $\nu\left(g\right)\in f_{g}^{-1}\left(0\right):\,\,\, f_{g}\left(\nu\right):=2^{3/2}\left[\Gamma\left(\frac{1-\nu}{2}\right)/\Gamma\left(-\frac{\nu}{2}\right)\right]+g$
and $U\left(a,b\right)$ denote the parabolic cylinder functions.
The symmetry operator $S_{+}:=S_{+}^{A}\otimes S_{+}^{B}$ serves
to compensate the symmetry breaking introduced in the Hamiltonian
\eqref{eq:Pair3}.

This solution gives a good approximation of the density patterns in
Fig. \ref{fig:Pair2} for intermediate to strong attractions $\left(\left|g_{AB}\right|<g_{\alpha}/2\right)$.
Also in the high coupling regime $\left(\left|g_{AB}\right|>g_{\alpha}/2\right)$
the model provides applicable predictions for the system's behavior.
Considering the molecule-molecule interaction term $\tilde{g}\delta\left(R_{1}\right)\equiv\left(g_{A}+g_{B}+2g_{AB}\right)\delta\left(R_{1}\right)$,
with large enough inter-species attraction $g_{AB}$ the effective
molecule-molecule interaction $\tilde{g}$ vanishes and even becomes
negative, i.e. attractive. That implies, more precisely, that for
$g_{AB}\approx g_{\alpha}/2$ a state forms where the $AB-$molecules
are condensed similar to a Bose-Einstein condensate (BEC). For further
increase of the interaction, the gas of $AB$-molecules collapses
and forms a bound state. Even though the introduced approximation
model gives reasonable predictions in that limit of very high inter-species
attractions $\left(\left|g_{AB}\right|>g_{\alpha}/2\right)$, it should
be handled with care, as the inter- and intra-species length scales
become comparable and therefore the scale separation breaks down (see
also Fig. \ref{fig:Pair2}).

These considerations can be supported by means of the quantities of
the one-body density matrix $\rho_{1}^{\left(\alpha\right)}\left(x,x'\right)=\frac{1}{N_{\alpha}}\hat{\left\langle \Psi\right.}_{\alpha}^{\dagger}\left(x\right)\hat{\Psi}_{\alpha}\left.\left(x'\right)\right\rangle $
and the \textit{pair density matrix}\[
\begin{array}{ccl}
\tilde{\rho}\left(x,x'\right) & := & \frac{1}{N_{A}N_{B}}\left\langle \Delta\right._{AB}^{\dagger}\left(x\right)\Delta_{AB}\left.\left(x'\right)\right\rangle \end{array},\]
with the {}``pair'' operator $\Delta_{AB}\left(x\right)\equiv\hat{\Psi}_{A}\left(x\right)\hat{\Psi}_{B}\left(x\right)$,
that annihilates an $AB$-pair {}``particle'' at the position $x$.
As $\Delta_{AB}\left(x\right)|\Psi\rangle$ is a {}``hole''-state,
i.e. a state where an $AB$-pair has been removed at the position
$x$, the pair density matrix embodies the overlap of two such {}``hole''-states.
The pair density matrix reflects the correlation inherent in the state
$\Psi$ between the positions $x$ and $x'$ on the level of $AB$-dimers,
as opposed to correlations of single particles $\alpha$ described
by $\rho_{1}^{\left(\alpha\right)}\left(x,x'\right)$ .

\begin{figure}
\includegraphics[width=0.35\columnwidth]{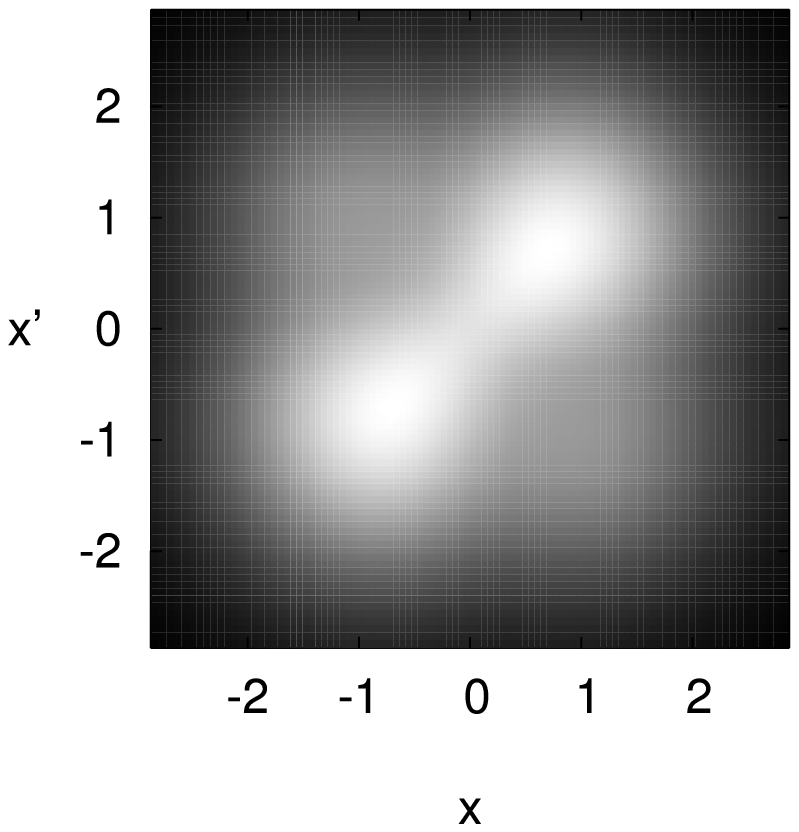}~~~~~~~~~~~~~\includegraphics[width=0.35\columnwidth]{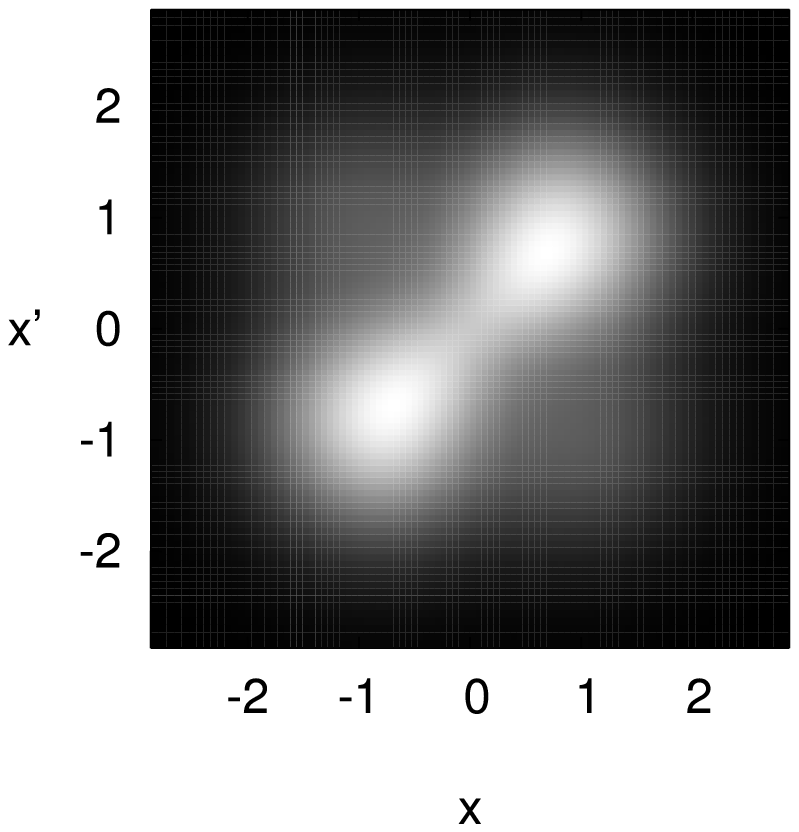}

\includegraphics[width=0.35\columnwidth]{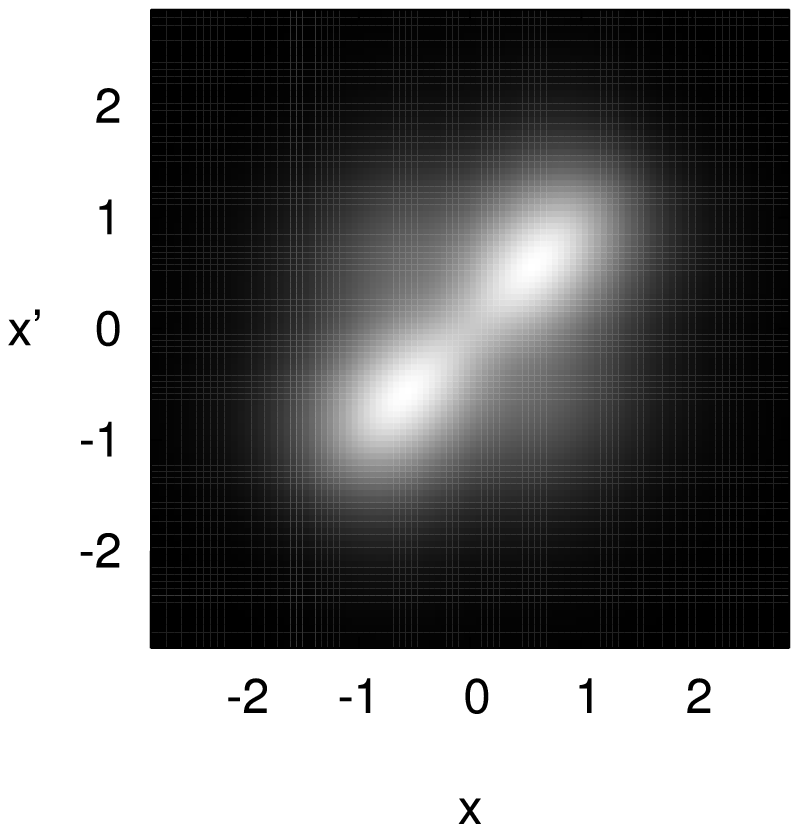}~~~~~~~~~~~~~\includegraphics[width=0.35\columnwidth]{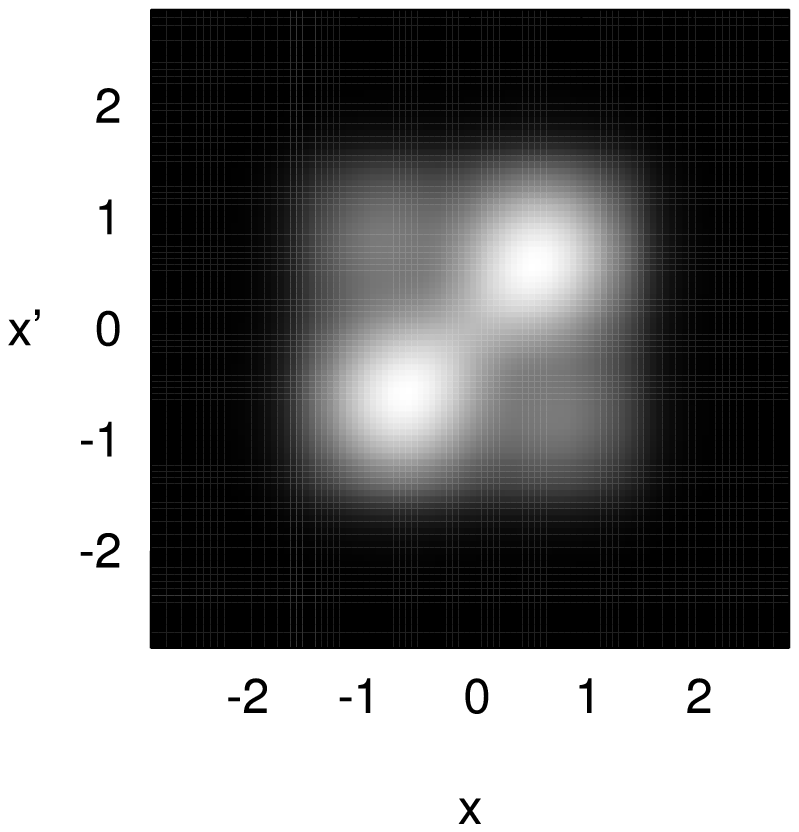}

\includegraphics[width=0.35\columnwidth]{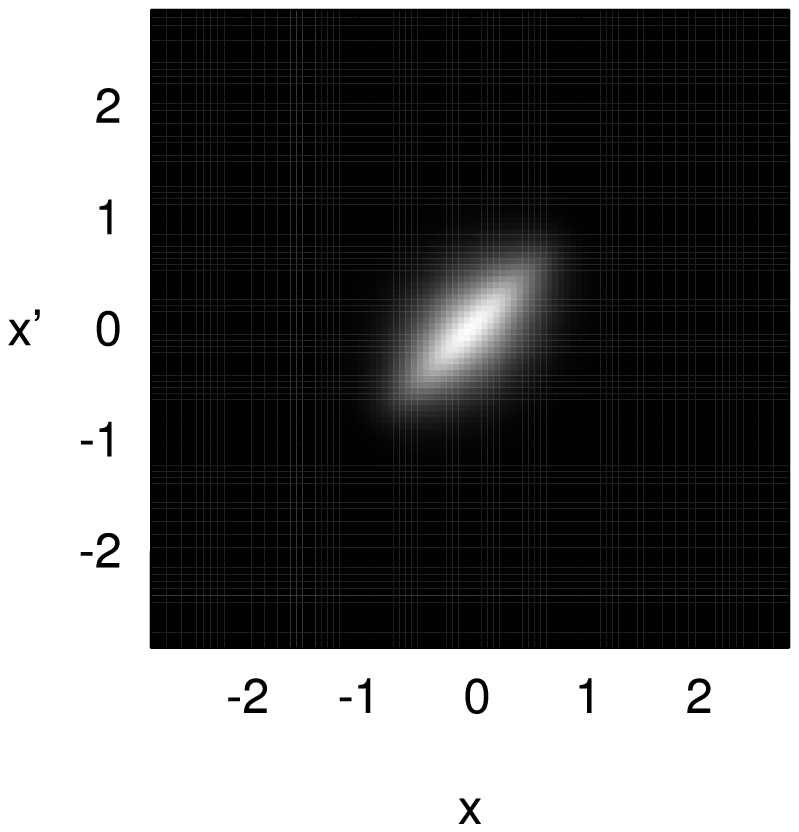}~~~~~~~~~~~~~\includegraphics[width=0.35\columnwidth,keepaspectratio]{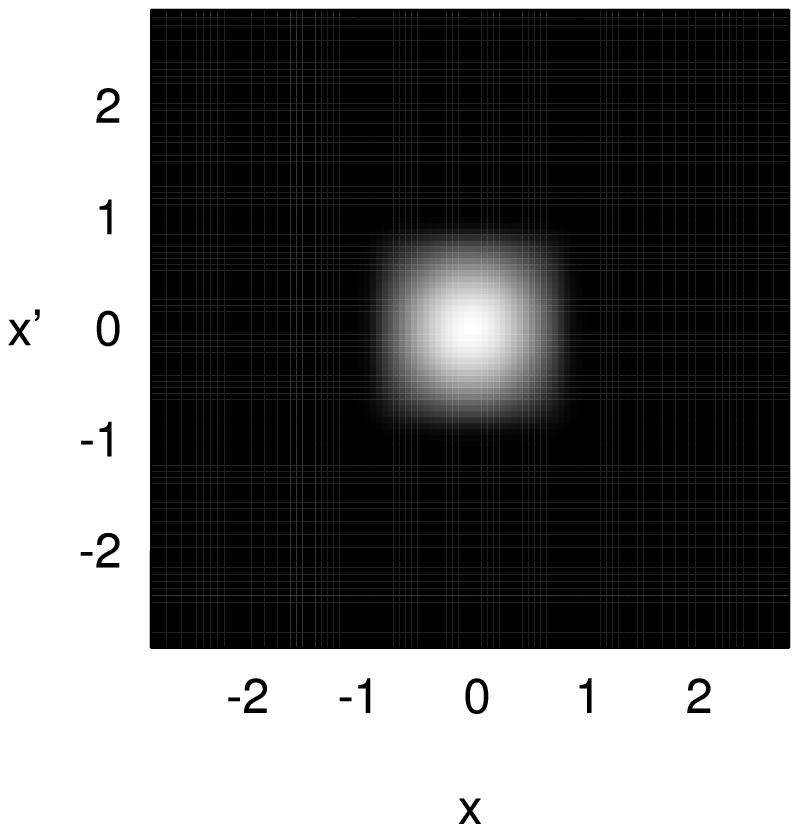}

\includegraphics[width=0.35\columnwidth]{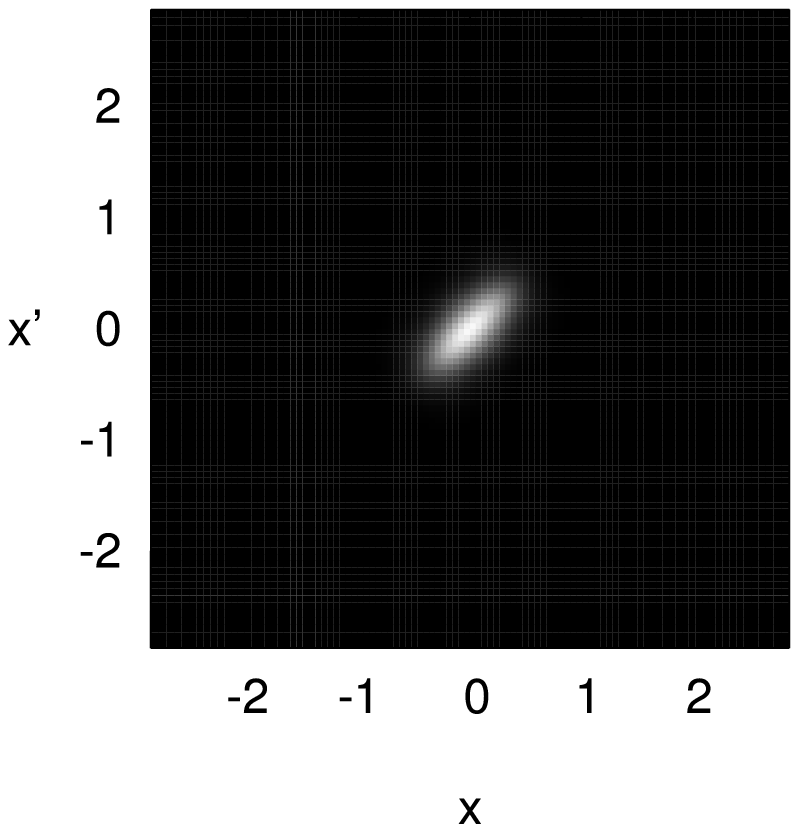}~~~~~~~~~~~~~\includegraphics[width=0.35\columnwidth]{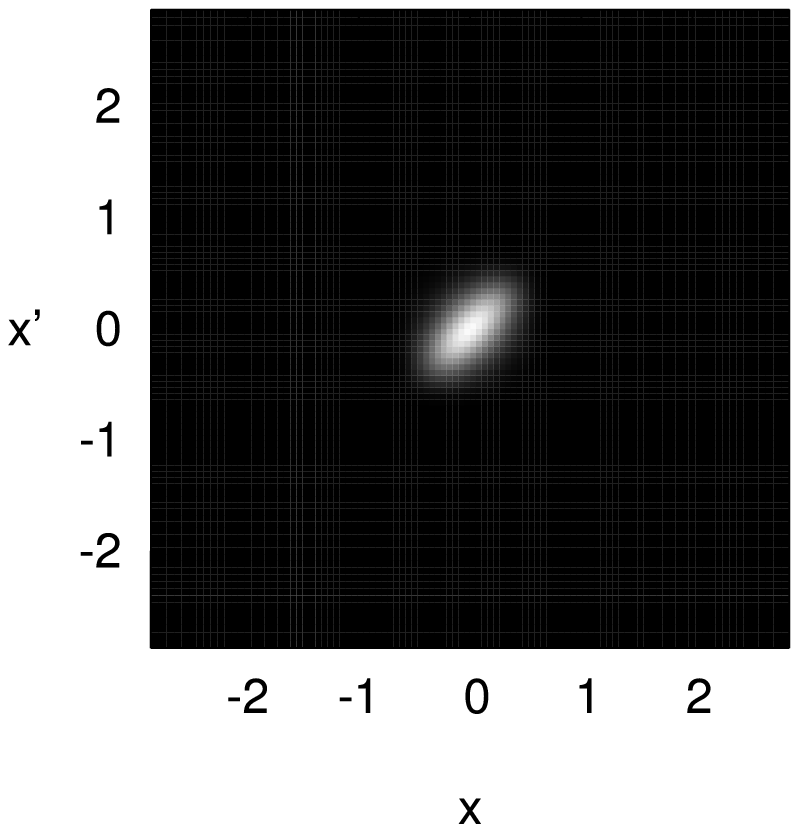}

\caption{One-body density matrix $\rho_{1}\left(x,x'\right)$ \textit{(left
column}) and pair density matrix $\tilde{\rho}\left(x,x'\right)$
(\textit{right column}) for inter-species couplings $g_{AB}=-0.001,\,-5.0,\,-20.0,\,-30.0$
(\textit{from top to bottom}).\label{fig:Pair3}}
\end{figure}

As shown in Fig. \ref{fig:Pair3}, the off-diagonal range of the pair
density matrix $\tilde{\rho}\left(x,x'\right)$ persists and even
slightly increases for the inter-species attractions up to $\left|g_{AB}\right|\approx5$,
where its appearance agrees well with the corresponding one-body density
matrix $\rho_{1}^{\left(M\right)}\left(x,x'\right)$ of identical,
\textit{fermionized} bosons with mass $M=2$ (in units of $m$).
This proves the existence of a paired state (MTG), as discussed above.
By contrast, the single-particle density matrix $\rho_{1}^{\left(\alpha\right)}\left(x,x'\right)$
(see Fig. \ref{fig:Pair3} left column) shows two peaks on the diagonal,
while the off-diagonal density steadily diminishes with increasing
$\left|g_{AB}\right|$. In this light a single $\alpha$-atom will
be in an incoherent superposition of left- (right-) localized states,
without any phase correlations. This has to be contrasted with the
phase correlations present for the pair density matrix (see Fig. \ref{fig:Pair3}
right column for $g_{AB}=-5.0$). Interestingly this may be compared
to a demixed state in the presence of repulsive inter-species interactions
\citep{zoellner08b}.

When further increasing the inter-species interaction to $\left|g_{AB}\right|\approx20$
the size of the system decreases; however, a seemingly perfect off-diagonal
long range order \citep{yang62} in the pair density matrix is attained,
which can be interpreted as (few-body analog of) a \textit{condensed
state} on the level of $AB$-molecules, $\tilde{\rho}\left(x,x'\right)=\varphi_{AB}^{*}\left(x\right)\cdot\varphi_{AB}$$\left(x'\right)$
(see Fig. \ref{fig:Pair3} right column for $g_{AB}=-20.0$). By contrast,
on the single-particle level, displayed in the one-body density matrix
$\rho_{1}^{\left(\alpha\right)}\left(x,x'\right)$, no condensed state
exists, but the two correlation peaks merge into one centered peak
concentrated on the diagonal $\left\{ x=x'\right\} $.

For inter-species interaction strength larger than the order of magnitude
of the intra-species interactions $\left|g_{AB}\right|\gg\left|g_{\alpha}\right|/2$
the system becomes highly bound beyond the $AB$-molecule level, as
is reflected in the decrease of the off-diagonal elements in the pair
density matrix $\tilde{\rho}\left(x,x'\right)$. This can be seen
as a \textit{collapse} from a molecular gas to a strongly interacting
cluster of $AB$-molecules. In contrast to a Bose-Fermi \citep{rizzi08}
and Fermi-Fermi mixture \citep{astrakharchik04b} the collapse in
a pure bosonic mixture is qualitatively different: In a Bose-Fermi
mixture only the bosons form a small region with high density, whereas
the fermions will be attracted up to a {}``Pauli-allowed'' density,
and Fermi-Fermi mixtures with $s$-wave interactions remain mechanically
stable even in the strongly attractive inter-species regime.%

Although, for definiteness, we restricted our discussion to a mixture
with particle numbers $N_{\alpha}=2$, the mechanism described by
Eq.~(\ref{eq:Pair3}) extends to the case of more molecules, $N_{\alpha}>2$.
For much larger systems $N\gg1$, of course, few-body effects as the
density oscillations seen in Fig.~\ref{fig:Pair1} will be smeared
out and the corresponding density profiles broadened due to repulsion,
as in the single-component Bose gas \citep{kolomeisky00}. Likewise,
the collapse witnessed for $g_{AB}\to-\infty$ will be much more pronounced,
as the center-of-mass width $\Delta R_{\mathrm{CM}}$ shrinks with
increasing atom number.

\subsubsection*{Fermionized attractive components}

\begin{figure}
\includegraphics[width=0.35\columnwidth]{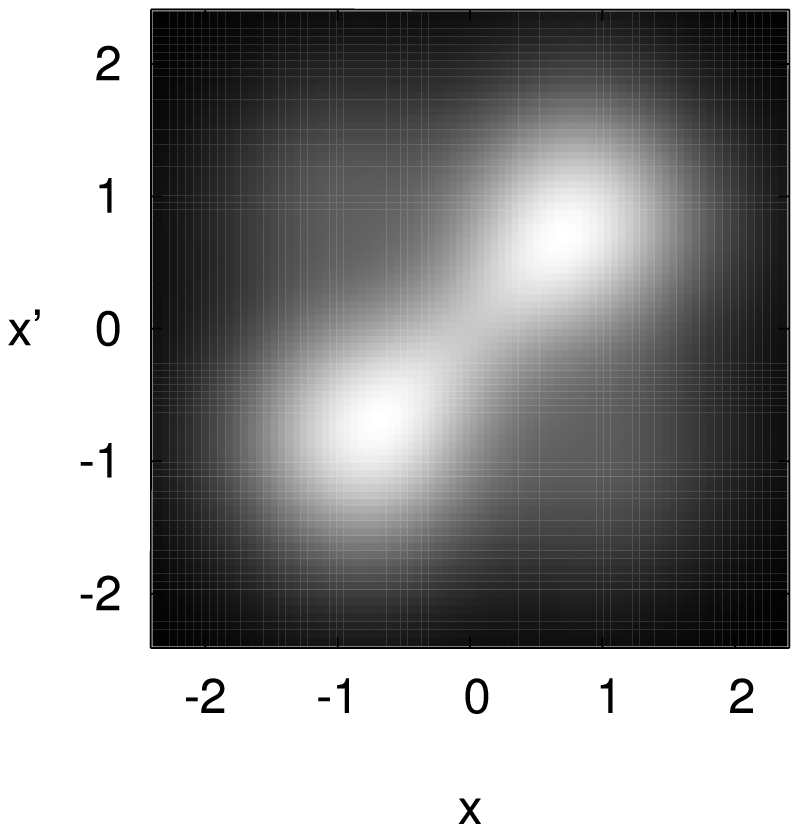}~~~~~~~~~~~~~\includegraphics[width=0.35\columnwidth]{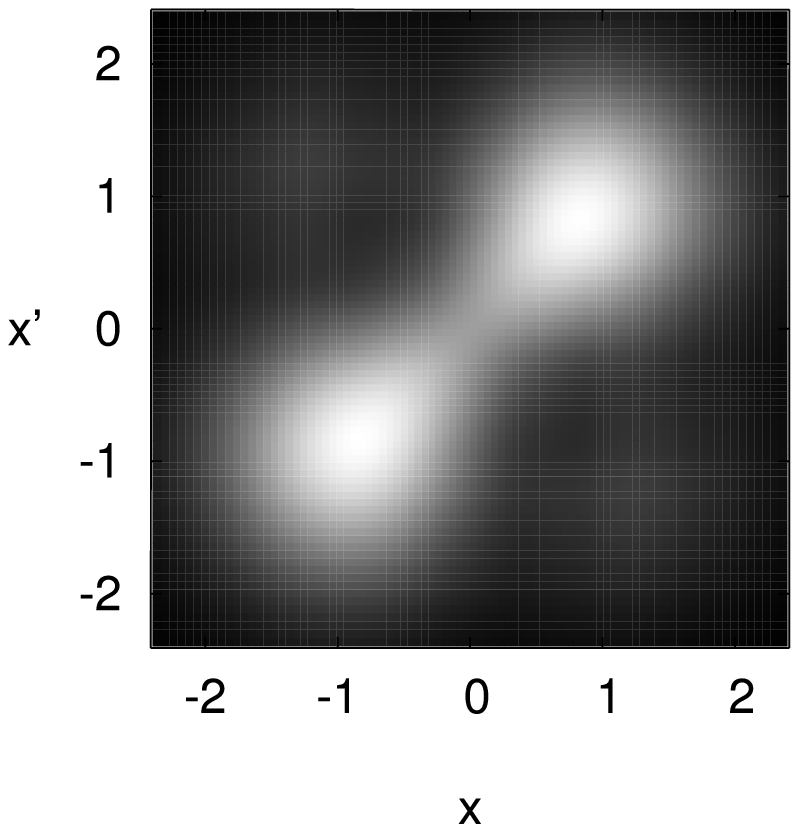}

\includegraphics[width=0.35\columnwidth]{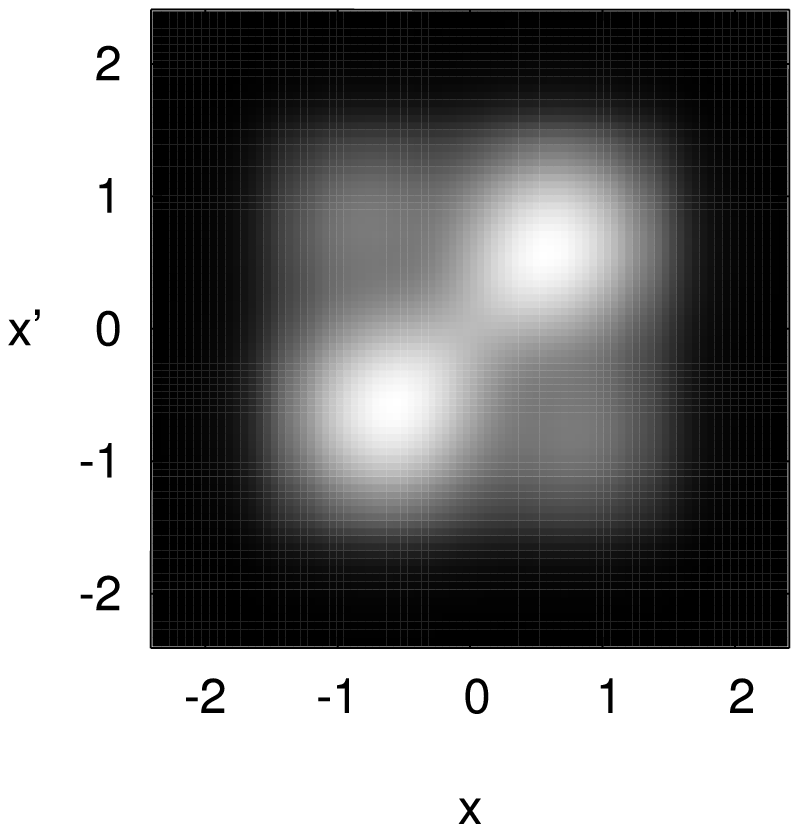}~~~~~~~~~~~~~\includegraphics[width=0.35\columnwidth]{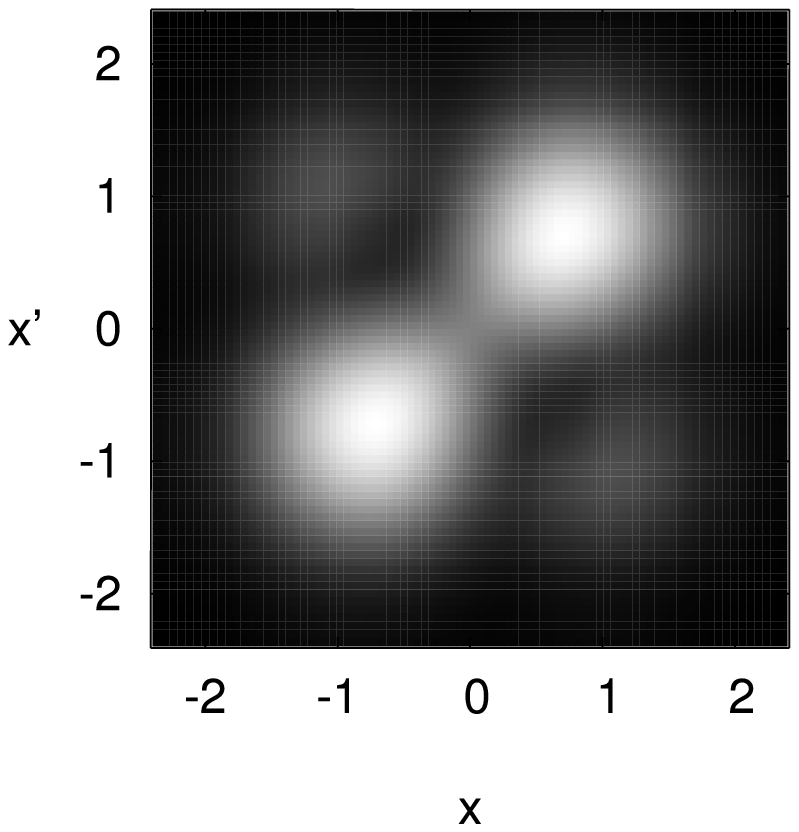}

\caption{Pair density matrix $\tilde{\rho}\left(x,x'\right)$ for \textit{$g_{\alpha}=25.0$}
(\textit{left column)} and the super-Tonks-Girardeau state \textit{}$g_{\alpha}=-15.0$
\textit{}(\textit{right column}) ($\alpha\in\left\{ A,B\right\} $).
The inter-species interaction parameters are $g_{AB}=-0.01,\,-5.0$
(\textit{from top to bottom}).\label{fig:Pair4}}
\end{figure}
It is known for identical bosons that fermionization can also be obtained
in the attractive interaction regime \citep{astrakharchik05,tempfli08},
where it is called the \textit{super-Tonks-Girardeau} state \textit{}(STG).
We show that the above pairing mechanism can also be observed in a
mixture with two attractively interacting, fermionized components
$\left(g_{\alpha}<0,\,\alpha\in\left\{ A,B\right\} \right)$. In this
case it is no longer the ground state but an excited state of the
system. We performed the numerical investigation for the exemplary
case of a mixture with $N_{\alpha}=2$ being situated in the energetically
lowest STG-state $\left(g_{\alpha}=-15.0\right)$. Direct comparison
with a system of repulsive, fermionized components shows the corresponding
process analogous to the formation of the molecular TG-gas, but with
smaller off-diagonal correlations (Fig.~\ref{fig:Pair4}). Clearly
the density profile of the STG-state is more localized in fragmented
regions than that of the TG-state (Fig. \ref{fig:Pair4}). The reason
is the finite intra-species interaction-strength (here $g_{\alpha}=-15.0$),
where the state is not completely fermionized. Since this quasi-STG-state
still has a non-vanishing, positive 1D-scattering length $a_{\alpha}=-\frac{2}{g_{\alpha}}>0$,
it is closer to a gas of spatially extended, hard-core particles (so-called
hard rods) than to a completely fermionized, point-like TG-gas \citep{astrakharchik05},
and localization effects are more pronounced, which can be observed
in the more profiled density.

\subsection{Weakly interacting components\label{sub:RepulsiveB}}

\begin{figure}
\includegraphics[width=0.7\linewidth]{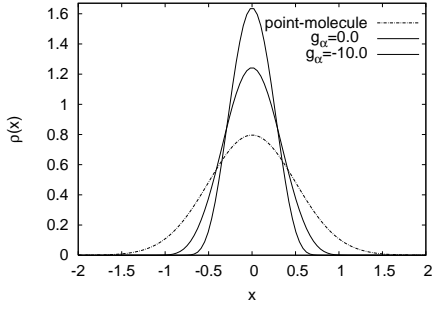}

\caption{One-body density $\rho\left(x\right)$ of $\frac{N}{2}=2$ point-like
molecules of mass $M=2$ with interaction strength $\tilde{g}=0$
(\textit{dashed}), two non-interacting components with $g_{AB}=-10.0$
(\textit{solid thin}), and two molecular components with $g_{AB}=-10.0$
(\textit{solid thick}) .\label{fig:Pair5}}
\end{figure}
Following the pathway to weak intra-species repulsion, the mechanism
of pair formation is getting constantly weaker till it vanishes in
the weak-interaction regime $\left(g_{\alpha}\sim1\right)$. In this
weakly interacting regime we turn to the limit case of two hardly
interacting, BEC-like components $\left(g_{\alpha}\approx0\right)$.
Compared to the case of two strongly repulsive components, there is
no formation of a condensed state of $AB$-molecules, but the system
collapses with the increase of the inter-species attraction. In other
words, between the $AB$-molecules there is always an effective attractive
interaction and thus for strong interaction a bright-soliton-like
state evolves. Figure \ref{fig:Pair5} displays a comparison of the
one-body densities of (\textit{i}) two identical point-molecules each
of mass $M=2$, which mirrors the case of very tightly bound, point-like
$AB$-molecules with no molecular interaction $\tilde{g}=0$, (\textit{ii})
an $N$-body bound state of the form \[
\begin{array}{l}
\Psi\left(X\right)\propto\\
{\,\Phi}_{0}\left(R\right)\left\{ e\right.^{-\frac{\left|g_{AB}\right|}{2}\left(\underset{i,j\leq2}{\sum}\left|x_{A_{i}}-x_{B_{j}}\right|\right)}\underset{\alpha\in\left\{ A,B\right\} }{\prod}e^{-\frac{\left|g_{\alpha}\right|}{2}\left|x_{\alpha_{1}}-x_{\alpha_{2}}\right|}\left.\right\} \end{array}\]
and (\textit{iii}) the case $g_{\alpha}=0$, which is a solitonic
state in between the two extremes. 

We note that the coherence between the {}``$AB$-molecules'' (as
evidenced in the pair density matrix) is slightly stronger compared
to the one-body level $\rho_{1}^{\left(\alpha\right)}\left(x,x'\right)$,
as there is just explicit interaction between the species ($g_{AB}$).
However, as there is no longer a scale separation, one cannot consider
this system simply as a gas of point-like molecules.

\subsection{Imbalanced components\label{sub:RepulsiveC}}

After having studied the mechanism for equal component settings, we
now want to highlight the effects of relaxing the equality of the
particle numbers and the intra-species interaction strengths.

\noindent %
\begin{figure}
\noindent \includegraphics[width=0.33\columnwidth]{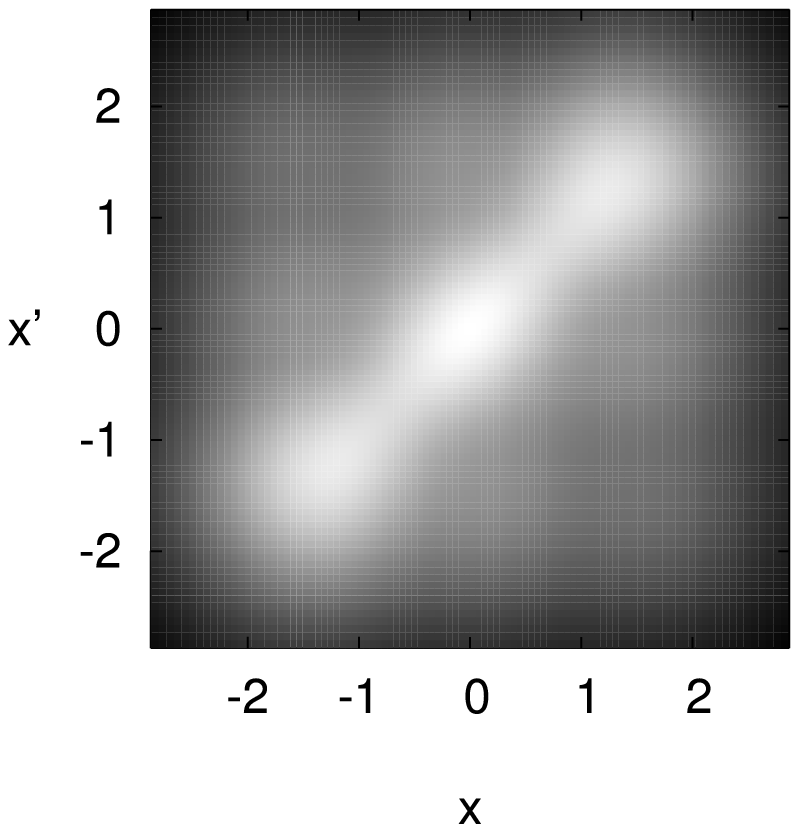}\includegraphics[width=0.33\columnwidth]{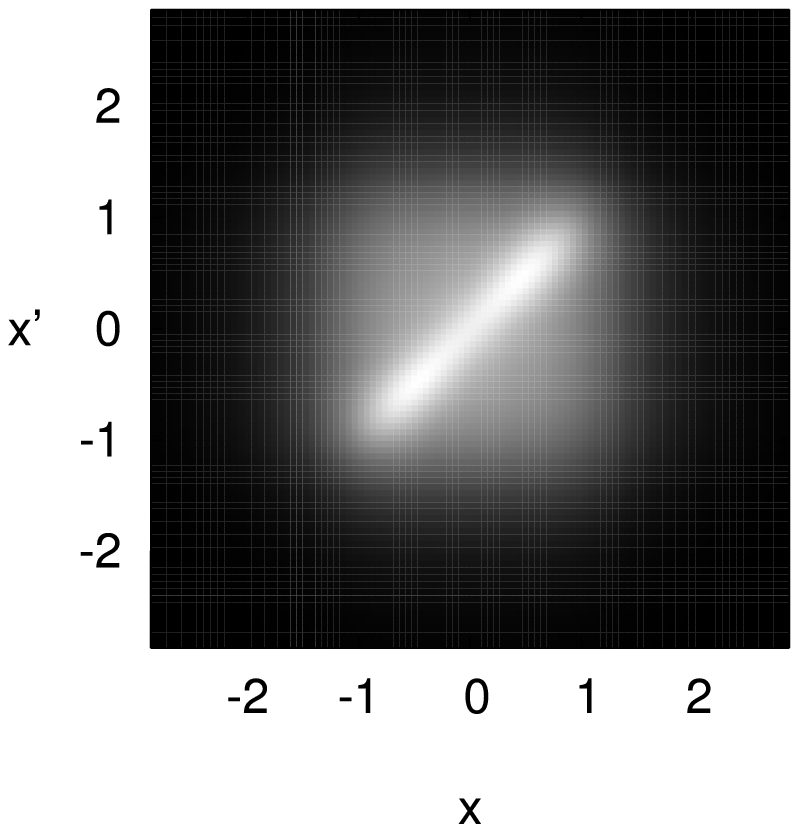}\includegraphics[width=0.33\linewidth]{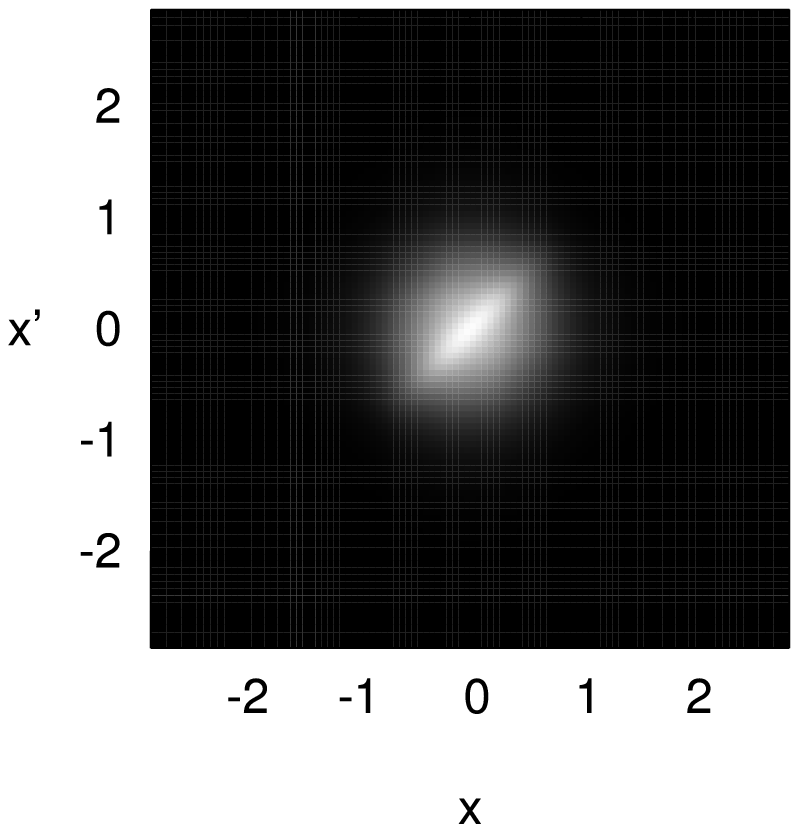}

\noindent \includegraphics[width=0.33\linewidth]{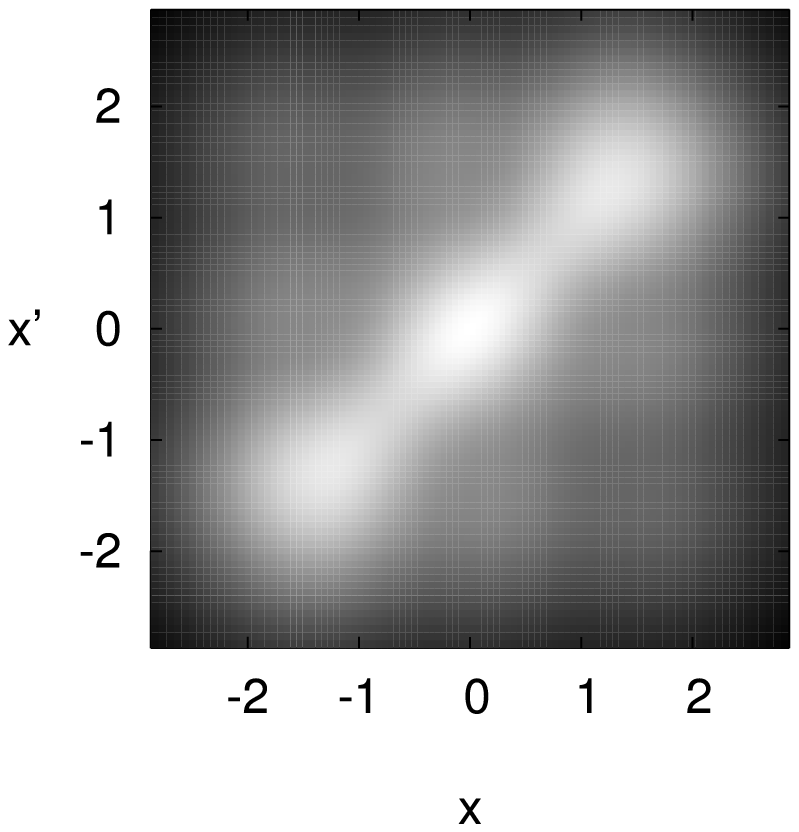}\includegraphics[width=0.33\linewidth]{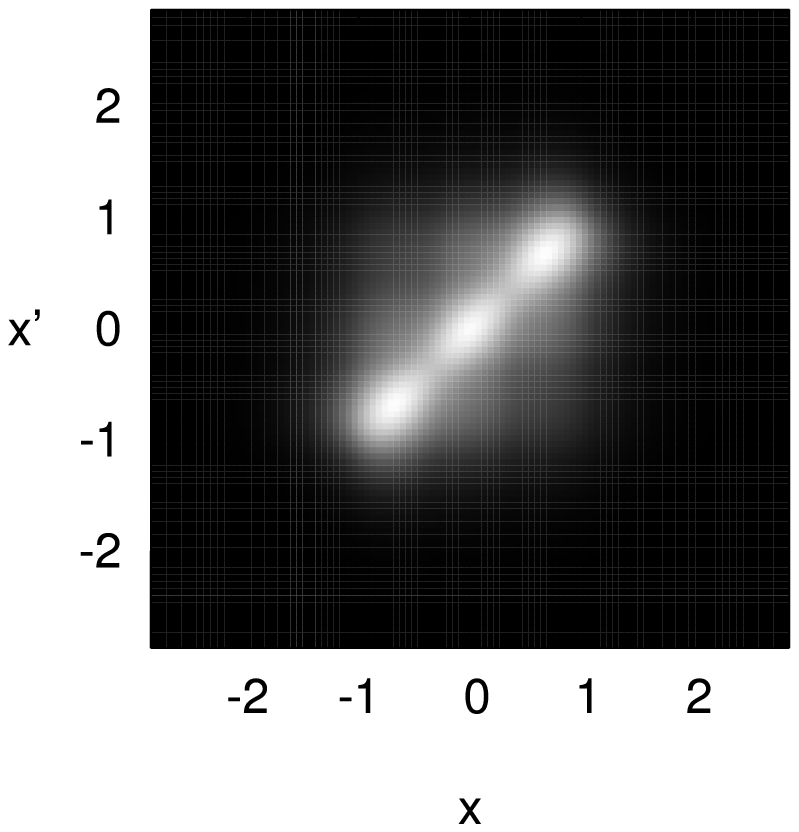}\includegraphics[width=0.33\columnwidth]{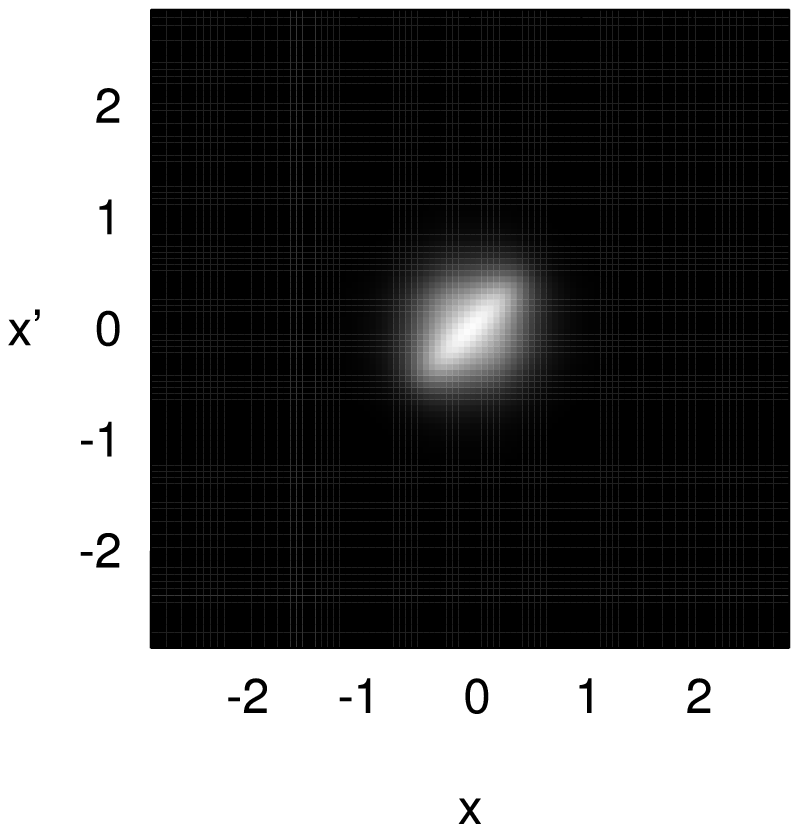}

\caption{One-body density matrix $\rho_{1}^{\left(A\right)}\left(x,x'\right)$
of two quasi-fermionized components ($g_{\alpha}=25.0$) for the inter-species
interaction strength $g_{AB}=-0.01,\,-10.0,\,-20.0$ (\textit{from
left to right}) of a mixture with particle numbers $N_{A}=3$ and
$N_{B}=2$ (\textit{upper row}), $N_{A}=N_{B}=3$ (\emph{l}\textit{ower
row}).\label{fig:Pair6}}
\end{figure}

\subsubsection*{Unequal particle numbers }

We first consider the case of unequal particle numbers $N_{A}\neq N_{B}$,
but still the same intra-species interaction strengths $\left(g_{A}=g_{B}\right)$.
We exemplify this on the case of two quasi-fermionized species ($g_{\alpha}=25.0$)
with particle numbers $N_{A}=3$ and $N_{B}=2$. On the way from weak
to very strong inter-species attractions $g_{AB}$, an analogous pathway
occurs as for fermionized binary mixtures with equal particle numbers,
as can be checked on the basis of the pair density matrix. That is,
a MTG state forms in the intermediate inter-species interaction regime,
followed by condensation and collapse for even higher inter-species
attractions. The effect of the difference in the particle numbers
(or particle densities) can be seen as a formation of two phases:
One consists of tightly bound $AB$-molecules, as in the case of equal
species numbers, and the other consists of $N_{d}\equiv\left|N_{A}-N_{B}\right|$
(here: $N_{d}=1$) {}``loosely bound'' spare particles, i.e. particles
that are hardly affected by the inter-species interaction (of course,
taking into account the proper particle exchange symmetries.) This
picture of loosely bound particles provides a good understanding of
the two-body density patterns in the intermediate to strong inter-species
interaction regime $\left(\left|g_{AB}\right|\lesssim10\right)$.
Furthermore this formation of, in this case, two $AB$-molecules and
one loosely bound particle manifests in the one-body density matrix
$\rho_{1}^{\left(A\right)}\left(x,x'\right)$ (Fig. \ref{fig:Pair6})
in the formation of two density peaks on the diagonal $\left\{ x'=x\right\} $
and the larger off-diagonal density compared to the balanced counterpart
$\left(N_{A}=N_{B}=3\right)$, respectively. This two-phase picture
breaks down as the system starts to collapse for larger attraction
(see Fig.~\ref{fig:Pair6}, $g_{AB}=-20$).

\subsubsection*{Unequal inter-species repulsions}

Now we consider unequal intra-species repulsions $g_{A}\neq g_{B}$.
For the sake of clarity, we keep the particle numbers equal $N_{A}=N_{B}$.%
\begin{figure}
\subfigure[]{\includegraphics[width=0.48\columnwidth]{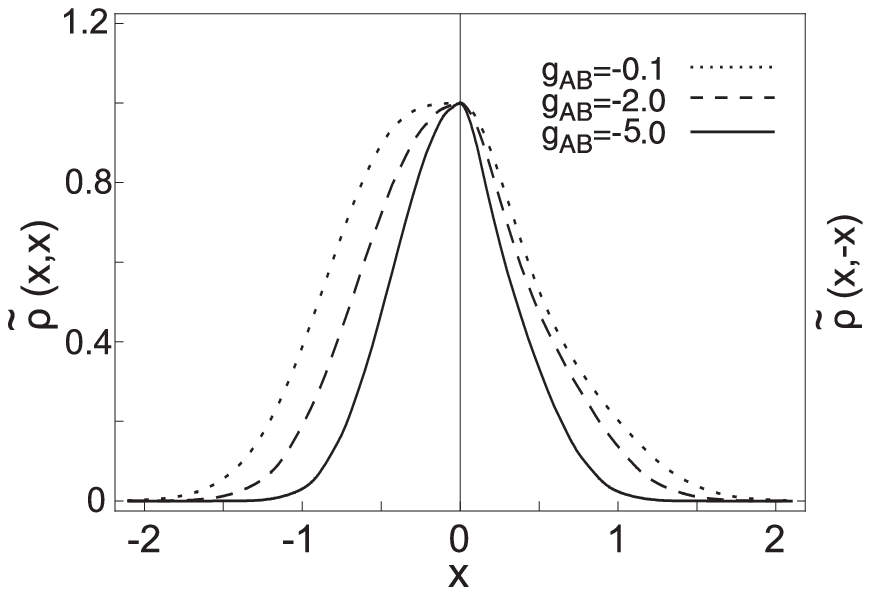}}\hfill{}\subfigure[]{\includegraphics[width=0.48\columnwidth]{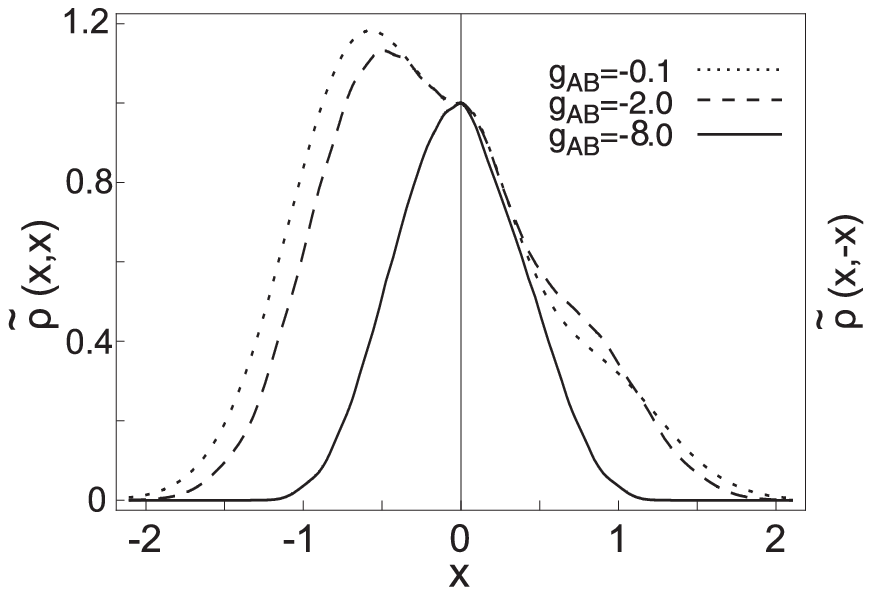}}

\subfigure[]{\includegraphics[width=0.48\columnwidth]{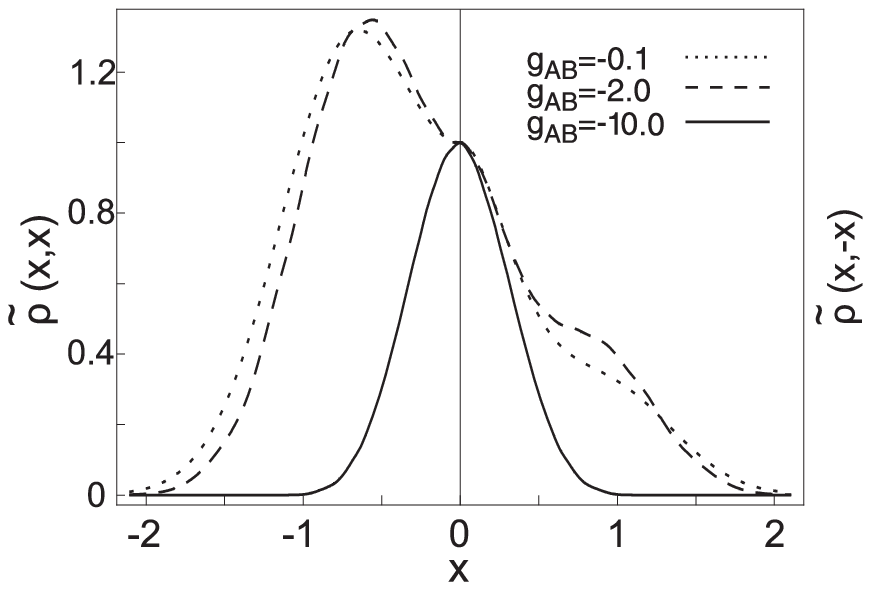}}\hfill{}\subfigure[]{\includegraphics[width=0.48\columnwidth]{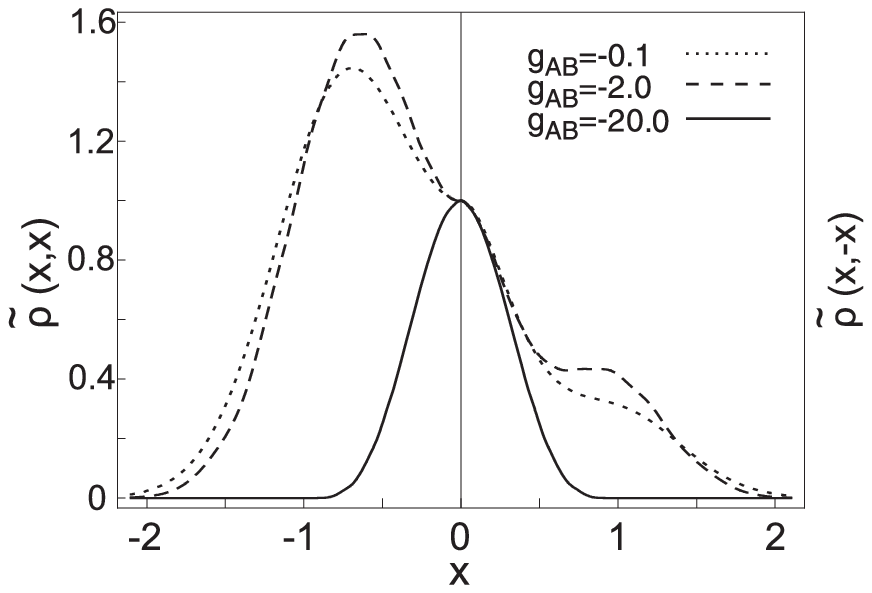}}

\caption{Pair density matrix $\tilde{\rho}\left(x,x'\right)$ for the particle
numbers $N_{A}=N_{B}=2$ plotted along the diagonal $\left\{ x'=x\right\} $
for $x\in\left[-2.2,0.0\right)$ and along the off-diagonal $\left\{ x'=-x\right\} $
for $x\in\left[0.0,2.2\right]$, of one fermionized component $g_{B}=25.0$
and intra-species interactions of the other component: (\textit{a})
$g_{A}=0.01$ (\textit{b}) $g_{A}=5.0$ (\textit{c}) $g_{A}=10.0$
(\textit{d}) $g_{A}=25.0$. The highest inter-species interaction
\textit{(solid line)} shows the best possible symmetry between on-
and off-diagonal. (The densities have been rescaled to the same maximal
value at the position $x=0$.)\label{fig:Pair7}}
\end{figure}
\begin{figure}
\includegraphics[width=0.7\columnwidth]{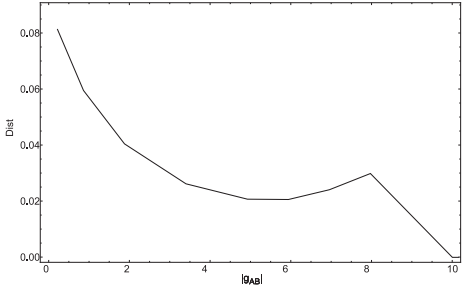}

\caption{Distance $\mathrm{Dist}$ between the one-body density maxima $\mathrm{Max}[\rho_{1}^{\left(\alpha\right)}\left(x,x'\right)]$
of the components $\alpha\in\left\{ A,B\right\} $ for different inter-species
interaction strength $g_{AB}$ and the intra-species interactions
$g_{A}=10.0$ and $g_{B}=25.0$. \label{fig:Pair8}}
\end{figure}
 We start with the case of all intra-species interactions corresponding
to the fermionization regime, here $g_{A}=10.0$ and $g_{B}=25.0$.
In the chosen example the species $A$ is weakly fermionized, but
still the system evolves similar to the case of two strongly fermionized
species discussed above, that is we observe the formation of a pronounced
Tonks-Girardeau pattern in the pair density matrix $\tilde{\rho}\left(x,x'\right)$,
indicating the MTG state. This is in line with Figure \ref{fig:Pair7},
which pictures the $\tilde{\rho}\left(x,x'\right)$-profile along
the diagonal $\left\{ x'=x\right\} $ and along the off-diagonal $\left\{ x'=-x\right\} $. 

During the formation of the MTG state, one can observe an assimilation
of, for instance, the one-body density matrices $\rho_{1}^{\left(A\right)}\left(x,x'\right)$
and $\rho_{1}^{\left(B\right)}\left(x,x'\right)$. The best overlap
is achieved about the value $\left|g_{AB}\right|\approx5.0$. To characterize
this increasing similarity, let us define the distance between the
peak positions $x^{\mathrm{max},\alpha}$ of each component's (diagonal)
density profile, i.e., where $\rho_{1}^{\left(\alpha\right)}\left(x^{\left(\mathrm{max},\alpha\right)},x^{\left(\mathrm{max},\alpha\right)}\right)$
is maximal). For for a given inter-species coupling $g_{AB}$, this
is denoted by $\mathrm{Dist}:=\left|\left|x^{\left(max,A\right)}\right|-\left|x^{\left(max,B\right)}\right|\right|$
(see Fig. \ref{fig:Pair8}). 

With a further increase of the inter-species interaction $(\left|g_{AB}\right|>5$),
the system collapses in a way characteristic for Bose-Fermi mixtures
\citep{rizzi08}. That is, the less repulsive component $A$ forms
a high density region in the center of the trap, whereas the strongly
fermionized component keeps its fermionic character up to higher inter-species
interactions $\left(\left|g_{AB}\right|\approx8.0\right)$. That decrease
of the density overlap is visualized in the light increase of the
distance $\mathrm{Dist}$ of the extrema of the one-body density for
$5<\left|g_{AB}\right|\lesssim8$ (Fig. \ref{fig:Pair8}). This characteristic
does not hold for very high inter-species interactions ($\left|g_{AB}\right|\gtrsim10$),
where the system collapses similarly to that of two equally fermionized
Bose components, discussed above. The comparison of the case at hand
(see Fig. \ref{fig:Pair7} (c)) with that of two fermionized components
(see Fig. \ref{fig:Pair7} (d)) shows that the length scale of the
off-diagonal profile is always smaller than that of the diagonal profile.
Thus no $g_{AB}$ exists for which a perfect off-diagonal long range
order is achieved in the pair density matrix $\tilde{\rho}\left(x,x'\right)$.
In this sense, the system starts to collapse, without forming a condensed
state on the $AB$-molecule level. Noteworthy, if both components
are equally weakly fermionized $\left(g_{A}=g_{B}=10.0\right)$, no
partial collapse occurs as in the above case, and a condensed state
can be observed.

Going towards components with intermediate and weak intra-species
repulsions (like $g_{A}\lesssim5$ shown in Fig. \ref{fig:Pair7}
(a-b)) the pair-formation is not visible anymore, and the system immediately
starts to collapse without forming a condensed state as in the case
before. In the extreme case of one {}``condensed'' component $\left(g_{A}\rightarrow0^{+}\right)$,
the approximation \eqref{eq:Pair3} above is not valid as the length
scales cannot be separated anymore. For unequal particle numbers,
the system can again be thought of as a two-phase system; hence if
the particle number of the fermionized state (say $B$) exceeds the
number of condensed particles, the coherence in the one-body density
$\rho_{1}^{\left(B\right)}\left(x,x'\right)$ has larger off-diagonal
elements due to spare (unbound) $B$-particles.

\section{Mixture with attractive components\label{sec:Attractive}}

In this section we complete our investigation by exploring mixtures
with one or more attractively interacting components ($g_{\alpha}<0$).

\subsection{Repulsive and attractive components\label{sub:AttractiveA}}

We start in the spirit of the above section with one component in
the fermionized interaction limit, i.e. $g_{B}=25.0$ and the other
in a bound state $g_{A}=-10.0$. The bound species are strongly localized
in the center of the trap and the feedback on that component is negligible
for any inter-species attraction. To explore this situation it is
natural to consider a simplified Hamiltonian, where the effect of
the localized species $A$ is replaced by an additional external potential
for the $B$ atoms $\delta U_{B}\left(x\right)=g_{AB}N_{A}\rho^{(A)}(x)$
\footnote{This mean-field ansatz is reminiscent of the case where species \emph{A}
is much heavier and may thus be treated as a classical potential for
\emph{B} \citep{zoellner08b}.%
}. Here we apply the analytically well studied split-trap $\rho^{(A)}(x)\approx\delta\left(x\right)$
\citep{Busch03,goold08},

\begin{equation}
\begin{array}{ccl}
\bar{H}_{B} & = & \overset{N_{B}}{\underset{i=1}{\sum}}\left(\frac{1}{2}p_{B_{i}}^{2}+\frac{1}{2}x_{B_{i}}^{2}+N_{A}g_{AB}\delta\left(x_{B_{i}}\right)\right)\\
 &  & +g_{B}\underset{i<j}{\overset{N_{B}}{\sum}}\delta\left(x_{B_{i}}-x_{B_{j}}\right).\end{array}\label{eq:Pair5}\end{equation}
Furthermore for $g_{B}\gg1$, one can map the fermionized component
on a \textit{non-interacting} fermionic system \citep{girardeau60}.
Consequently one obtains for the exemplary case of $N_{B}=2$ $B$-particles
the simple solution $\Psi_{0}^{B}=\left|\Psi_{0}^{Fermion}\right|=\frac{1}{\sqrt{2}}\left|\Phi_{0}\left(x_{1}\right)\Phi_{1}\left(x_{2}\right)-\Phi_{0}\left(x_{2}\right)\Phi_{1}\left(x_{1}\right)\right|$,
where $\Phi_{i}$ denotes the $i$-th single-particle eigenstate of
a split-trap. 

The validity of the approximation becomes rapidly better with increasing
number of particles ($N_{A}$) in the bound state, as the width of
one-body density $\rho^{\left(A\right)}\left(x\right)$ scales as
$\frac{1}{\sqrt{N_{A}}}$, and hence converges towards a $\delta$-type
potential in the limit of large particle numbers $\left(N_{A}\rightarrow\infty\right)$.
The agreement is astonishingly good already with relatively few particles
$N_{A}\geq4$.%
\begin{figure}
\includegraphics[width=0.33\linewidth]{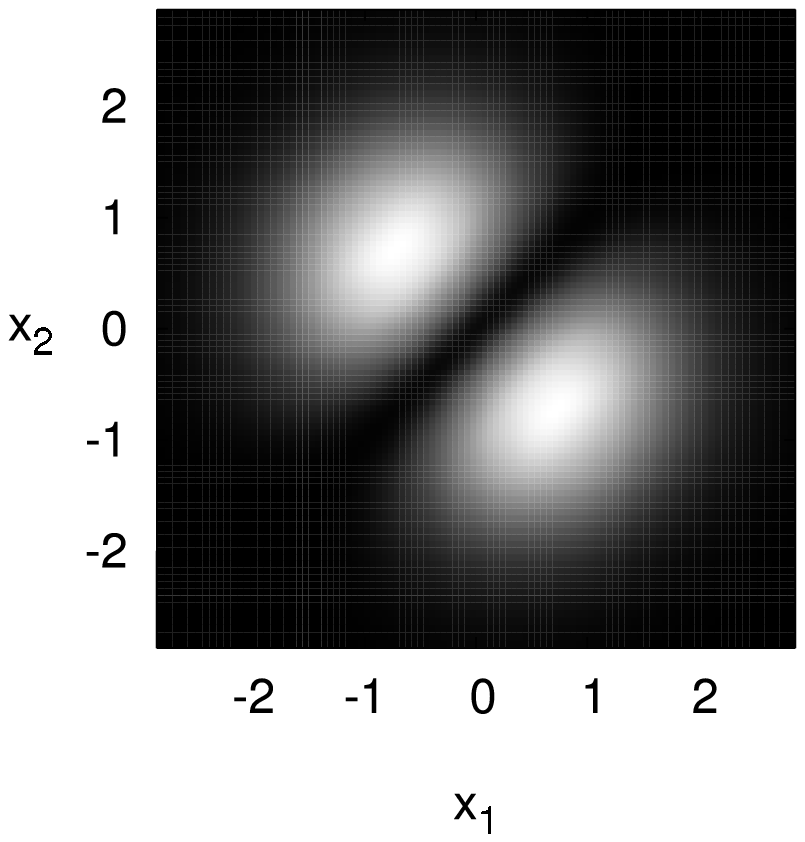}\includegraphics[width=0.33\linewidth]{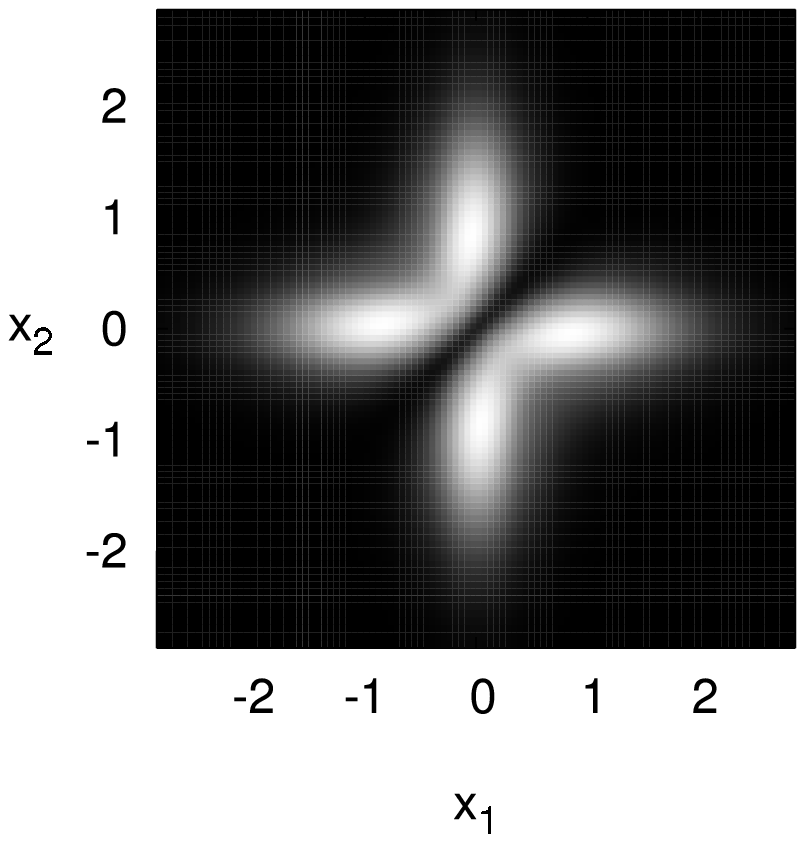}\includegraphics[width=0.33\linewidth]{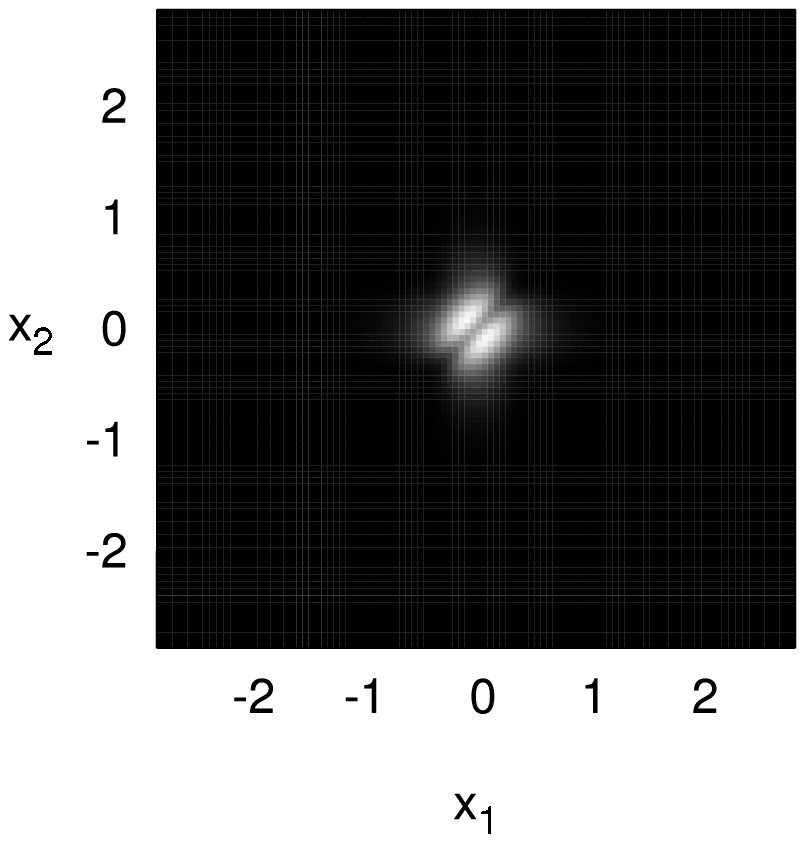}

\caption{Two-body correlation function $\rho_{BB}\left(x_{1},x_{2}\right)$
of a mixture with one molecular species $g_{A}=-10.0$ ($N_{A}=4$)
and one species of $N_{B}=2$ repulsive bosons, $g_{B}=25.0$, for
the inter-species interaction parameter $g_{AB}=-0.01,\,-1.0,\,-5.0$
\textit{(from left to right)}.\label{fig:Pair9}}
\end{figure}
 In Fig. \ref{fig:Pair9}, the exact two-body correlation function
$\rho_{BB}\left(x_{1},x_{2}\right)$ for $N_{B}=2$ particles is shown.
As it turns out, the picture of a non-interacting fermionic system
applies very well up to intermediate inter-species interactions $\left|g_{AB}\right|<2$
(in the case of $N_{A}=4$). This predicts that if for intermediate
inter-species interaction $g_{AB}\approx-1$ one detects a $B$-particle
aside the trap center $\left(x_{B_{1}}\approx\pm1\right)$, the other
$B$-particle is located at the center of the trap $\left(x_{B_{2}}\approx0\right)$.
(For $N_{B}>2$ an additional density contribution emerges on the
off-diagonal.)

\begin{figure}
\includegraphics[width=0.7\columnwidth]{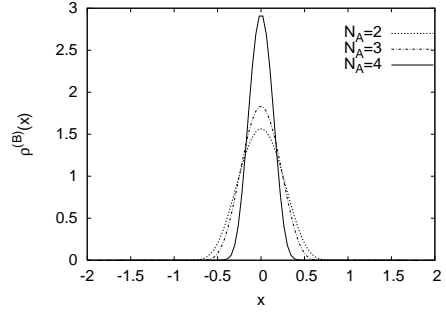}

\caption{One-body density $\rho^{\left(B\right)}\left(x\right)$ of $N_{B}=2$
non-interacting $B$-particles $\left(g_{B}=0.0\right)$, with different
$A$-particle numbers $N_{A}$ \textit{}in a bound state $\left(g_{A}=-10.0\right)$
and the inter-species interaction strength $g_{AB}=-10.0$.\label{fig:Pair10}}
\end{figure}

Whereas for the model $\delta$-type potential the two-body density
$\rho_{BB}\left(x_{1},x_{2}\right)$ would remain in the (increasingly
sharp) cross-shaped pattern even for $g_{AB}\to-\infty$, this is
not the case for the system at hand (see Fig. \ref{fig:Pair9} for
$g_{AB}=-5.0$). Due to the nonzero width of the additional potential
caused by the $A$-particles, for high enough inter-species attraction
all of the B particles can be accommodated in the {}``$A$-potential''
as a whole (unlike for a $\delta$-type potential). That is illustrated
in Fig. \ref{fig:Pair9} for $g_{AB}=-5.0$, where again the familiar
TG-density pattern can be observed, but with the spatial extension
of the $A$-particle density (compare with Fig. \ref{fig:Pair10}).
This behavior in the high interaction regime can also be observed
for higher $B$-particle numbers. 

\subsection{Two attractive components\label{sub:AttractiveB}}

Extending the results of the last section, we start with the case
of one weakly interacting, i.e., {}``condensed'' component $\left(g_{B}\approx0\right)$,
and the other component again in a bound state $\left(g_{A}=-10.0\right)$.
We can again apply the previous split-trap approximation on a system
with $N_{B}=2$, i.e. the condensed $B$-particle feel an effective
short-range potential at the center of the harmonic trap. Again with
increasing particle-number $N_{A}$ in the molecular state, the approximation
is getting better. However, for the length scales of the two components
to differ as distinctly as in the case of a fermionized component,
the agreement with the split-trap model above requires more $B$-particles
in the bound state. If we assume as a model a condensed state in component
$B$ $\left(g_{B}=0\right)$ and a tightly bound, $\delta$-type state
in component $A$, the model Hamiltonian \eqref{eq:Pair5} reduces
to

\begin{equation}
\begin{array}{ccl}
\bar{H}_{B} & = & \overset{N_{B}}{\underset{i=1}{\sum}}\left(\frac{1}{2}p_{B_{i}}^{2}+\frac{1}{2}x_{B_{i}}^{2}{+N_{A}g}_{AB}\delta\left(x_{B_{i}}\right)\right)\end{array},\label{eq:Pair6}\end{equation}
with the solution \citep{goold08} $\Psi_{0}\left(X_{B}\right)\propto\exp\left(-\frac{\left(x_{B_{1}}^{2}+x_{B_{2}}^{2}\right)}{2}\right)$$\prod_{i=1}^{N_{B}}U\left(\frac{1}{2}-\frac{E_{0}}{2},\frac{1}{2},x_{B_{i}}^{2}\right)$,
which evolves with increasing inter-species attraction towards a state
analogous to the bound state of the $\delta$-potential.

For higher attractive interactions also in component $B$, the model
is not applicable anymore. In the limit of highly attractive components,
with increasing inter-species interaction strength the system forms
an entire bound state. For the special case of equal interaction strength
$g_{\alpha}=g_{AB}\equiv g$, one can map the system to a bound state
of $N_{A}+N_{B}$ identical particles $\Psi_{0}\left(X\equiv\left(X_{A},X_{B}\right)\right)\propto\Phi_{0}\left(R\right)\exp\left(-\frac{\left|g\right|}{2}\underset{i<j}{\sum}\left|x_{i}-x_{j}\right|\right)$.

\section{Summary}

In conclusion, we have investigated the ground state of a two-component
Bose gas in a one-dimensional harmonic trap throughout the crossover
from weak to strong inter-component \textit{attraction}. We have highlighted
different pathways depending on the choices of the different intra-component
couplings, and indicated how they can be understood in terms of simplified
models. For two quasi-fermionized components (i.e. Tonks-Girardeau
states), the system forms a molecular Tonks-Girardeau gas in the intermediate
inter-component interaction regime, which consists of bound pairs
containing one particle of each component. In the strongly attractive
regime, we demonstrated the condensation of the bound pairs, followed
by the collapse of the system beyond a critical attraction. We showed
the analogous mechanism for attractively fermionized components, that
is components in the super-Tonks-Girardeau regime. Relaxation of the
condition of two equally fermionized components leads to a modified
pathway: In the case of just one fermionized component, the formation
of a molecular Tonks-Girardeau gas can still be observed for high
enough repulsion within the second component, but the collapse occurs
in analogy to Bose-Fermi mixtures without condensation of the bound
pairs, in contrast to the case of comparable repulsions. In the regime
of intermediate (inter-species) attraction, unequal (number-) densities
in the components have been found. These can be understood as two
phases, one consisting of molecular pairs of each component, and the
other phase consisting of loosely bound particles. For mixtures with
one strongly attractive component, we showed that this component can
be interpreted as an additional external $\delta$-function potential
for the other component, in case both length scales can be well separated.
The investigation of these intriguing pairing scenarios paves the
way toward studying their quantum dynamics, such as the tunneling
of molecular pairs in multi-well traps.

\begin{acknowledgments}
Financial support from the Landesstiftung Baden-Württemberg through
the project {}``Mesoscopics and atom optics of small ensembles of
ultracold atoms'' is gratefully acknowledged by P.S. and S.Z. We
thank H.-D. Meyer, X.-W. Guan, and O. Alon \textbf{}for fruitful
discussions.
\end{acknowledgments}
\bibliographystyle{/home/nbia/zoellner/bin/prsty}
\bibliography{fermi,mctdh,phd}

\end{document}